\documentclass[showpacs,floatfix,superscriptaddress]{revtex4}
\usepackage{graphicx}
\usepackage{multirow}
\usepackage{tabularray}
\usepackage{epsfig}
\usepackage{bm}
\usepackage[utf8]{inputenc}
\usepackage{amssymb}
\usepackage{float}
\usepackage{amsmath}
\usepackage{dcolumn}
\usepackage{latexsym}
\usepackage{subfig}
\usepackage{amsthm}
\usepackage{array}
\usepackage{framed}
\usepackage{cancel}
\usepackage[colorlinks]{hyperref}
\usepackage[usenames,dvipsnames]{color}
\hypersetup{
     breaklinks=true,
    pdfstartview={FitH},    
    colorlinks=true,       
    linkcolor=blue,          
    citecolor=red,        
    filecolor=magenta,      
    urlcolor=blue,           
    anchorcolor=green,      
    linktocpage=true
}

\def\doi{http://doi.org}

\begin{document}

\title{Thin Accretion Disks around Rotating Charged Black Holes in an Effective Higher-Curvature Spacetime}

\author{Mohammad Hassani}
\email[]{Mohammad.hassanei@gmail.com}
\affiliation{Department of Theoretical Physics, Faculty of Science, University of Mazandaran,\\
P. O. Box 47416-95447, Babolsar, Iran}

\author{Kourosh Nozari}
\email[]{knozari@umz.ac.ir (Corresponding Author)}
\affiliation{Department of Theoretical Physics, Faculty of Science, University of Mazandaran,\\
P. O. Box 47416-95447, Babolsar, Iran}
\author{Sara Saghafi}
\email[]{s.saghafi@umz.ac.ir}
\affiliation{Department of Theoretical Physics, Faculty of Science, University of Mazandaran,\\
P. O. Box 47416-95447, Babolsar, Iran}
\affiliation{School of Physics, Damghan University, Damghan.\\
 3671645667, Iran}

\begin{abstract}
We investigate the structure and emission properties of a thin accretion disk around a rotating charged black hole described by an effective higher-curvature--inspired spacetime, constructed as a phenomenological deformation of the Kerr--Newman geometry. In this framework, the deformation is introduced through a modification of the metric function \( \Delta \) by an effective Gauss--Bonnet--like parameter \( \alpha \), such that the spacetime reduces to the standard Kerr--Newman solution in the limit \( \alpha \to 0 \). Adopting a kinematical approach, we use test-particle motion to derive the specific energy, specific angular momentum, and angular velocity of circular orbits, and analyze the effects of the parameters \( \alpha \) and charge \( Q \) on the innermost stable circular orbit (ISCO), radiative efficiency, radiation flux, temperature, and differential luminosity of the disk. We find that increasing \( \alpha \) shifts the ISCO inward and enhances the disk’s radiation flux and temperature, while the presence of charge suppresses these quantities due to electrostatic effects. Our results demonstrate that effective higher-curvature deformations of rotating black hole spacetimes can lead to observable deviations from the Kerr case, highlighting accretion disks as sensitive probes of strong-gravity effects without relying on a specific underlying gravitational theory.

\vspace{12 pt}

\end{abstract}

\maketitle

\enlargethispage{\baselineskip}
\tableofcontents

\section{Introduction}\label{1}

The scientific study of the Universe has evolved from early descriptive models of the cosmos, such as Ptolemy’s geocentric picture, to modern relativistic descriptions founded on Albert Einstein’s theory of General Relativity (GR) \cite{1}. Among the most striking predictions of GR are black holes—regions of spacetime characterized by extreme curvature and causal structure. In four spacetime dimensions, the Schwarzschild, Kerr, and Kerr–Newman solutions provide the classical description of non-rotating, rotating, and rotating charged black holes, respectively, revealing fundamental features such as event horizons, singularities, and ergospheres \cite{2,3}.

Despite its remarkable success in describing gravitational phenomena from laboratory to astrophysical scales, GR faces well-known conceptual and phenomenological challenges. At small scales, GR is incompatible with quantum mechanics, while at large scales it requires the introduction of dark matter and dark energy components to account for galactic rotation curves and the late-time accelerated expansion of the Universe \cite{4}. These issues have motivated extensive efforts to explore extensions and modifications of GR, particularly those incorporating additional geometric or curvature-related effects \cite{5}.

Among such approaches, theories involving higher-curvature corrections—most notably Gauss–Bonnet–type terms—have attracted significant attention. The Gauss–Bonnet invariant arises naturally in low-energy effective actions of string theory and leads to second-order field equations in higher dimensions \cite{6,7}. In four dimensions, however, the Gauss–Bonnet term is purely topological and does not contribute dynamically unless additional degrees of freedom are introduced. Several proposals have been put forward to construct four-dimensional frameworks exhibiting Gauss–Bonnet–like effects \cite{7,08,008}, and a variety of black hole geometries inspired by these constructions have been studied \cite{8,9,10,11}. At the same time, it has become clear that consistent realizations generally involve scalar–tensor dynamics and that rotating solutions cannot, in general, be generated by the Newman–Janis algorithm. These conceptual subtleties necessitate a careful interpretation of rotating Gauss–Bonnet–inspired black hole metrics.

In light of these developments, it is both natural and useful to adopt a phenomenological perspective, in which rotating black hole geometries incorporating higher-curvature–inspired deformations are treated as effective spacetimes, without assuming that they arise as exact vacuum solutions of a specific four-dimensional gravitational action. Such an approach has proven valuable in exploring how deviations from the Kerr or Kerr–Newman geometry may manifest themselves in strong-gravity observables, particularly in astrophysical environments where precise measurements are becoming increasingly feasible \cite{47}.

Rotating charged black holes described by effective metrics of this type may be viewed as deformations of the Kerr–Newman spacetime, characterized by the mass 
$M$, rotation parameter $a$, charge parameter $Q$, and an additional parameter $\alpha$ encoding higher-curvature-inspired corrections. The influence of these parameters on spacetime geometry—through modifications of horizons, ergospheres, and geodesic structure—provides a powerful framework for studying strong-field gravitational phenomena and their observational signatures \cite{12,13,14,15}. Importantly, such studies do not rely on the validity of a particular fundamental theory, but instead focus on the kinematical consequences of deviations from the Kerr geometry.

One of the most important astrophysical processes governed by strong gravitational fields is accretion. Accretion disks around black holes consist of plasma and hot gas rotating in the equatorial plane and gradually spiraling inward under the influence of gravity. During this process, gravitational potential energy is converted into heat and electromagnetic radiation, making accretion disks among the most luminous astrophysical systems in the Universe \cite{40}. Accretion phenomena play a central role in powering active galactic nuclei (AGNs) and X-ray binaries \cite{22}, and they provide a direct observational window into the geometry of spacetime near black holes.

It is important to note that, in realistic astrophysical environments, black holes are expected to carry negligible net electric charge due to efficient neutralization by surrounding plasma. Therefore, the inclusion of a charge parameter in the present work should be interpreted primarily as a theoretical and phenomenological extension rather than a direct astrophysical requirement. In this context, the charge parameter $Q$ serves as a useful tool for exploring deviations from the Kerr geometry and for facilitating comparisons with the Kerr--Newman spacetime, which represents the most general stationary black hole solution in General Relativity. Moreover, effective charge-like terms may arise in certain modified gravity scenarios or in the presence of electromagnetic fields, allowing $Q$ to act as a proxy for additional interactions. Consequently, our analysis focuses on the qualitative impact of such parameters on accretion disk properties, rather than on the existence of highly charged astrophysical black holes.

The theoretical foundations of accretion disk models were established through seminal works by Hoyle and Lyttleton \cite{69}, Prendergast and Burbidge \cite{70}, and Shakura and Sunyaev \cite{22}, who introduced the thin-disk model with turbulent viscosity. Page and Thorne later extended this framework to rotating black holes, providing a relativistic description of thin accretion disks in stationary, axisymmetric spacetimes \cite{44}. Since then, accretion disks have been extensively studied using analytical models, numerical simulations, and magnetohydrodynamic approaches \cite{23,71,72,73,74,75,76,77,78,111,112,113,114,115,116,117,118}. Recent observational advances, particularly through the Event Horizon Telescope (EHT) and X-ray observatories such as NuSTAR, have enabled direct imaging and spectral studies of black hole environments \cite{79,80,13}, further motivating detailed theoretical investigations.

A key quantity governing the structure and emission properties of accretion disks is the innermost stable circular orbit (ISCO). The location of the ISCO determines the inner edge of the disk and strongly influences the radiative efficiency, spectral properties, and luminosity of the emitted radiation \cite{23}. The ISCO radius depends sensitively on the spacetime geometry and is therefore an excellent probe of deviations from the Kerr solution. Modifications of the spacetime geometry induced by effective higher-curvature parameters or charge can alter the specific energy, angular momentum, and angular velocity of particles at the ISCO, thereby affecting accretion efficiency and observable disk properties \cite{24,10,26}.

Another hallmark of rotating spacetimes is the frame-dragging effect, whereby the rotation of the central object induces a twisting of spacetime that affects particle motion and inertial frames \cite{27}. Frame dragging plays a crucial role in determining orbital dynamics, ergosphere structure, and energy extraction processes, and has been experimentally confirmed in the weak-field regime by missions such as Gravity Probe B and LAGEOS/LARES \cite{28}. In effective rotating spacetimes incorporating higher-curvature–inspired deformations, frame-dragging effects may differ quantitatively from those of the Kerr geometry, leading to potentially observable consequences in accretion disk dynamics \cite{29,12}.

In addition to circular orbits and frame dragging, the study of equipotential and von Zeipel surfaces provides valuable insight into rotational effects in strong gravitational fields. These surfaces play an important role in understanding disk stability, fluid equilibrium, and the redistribution of angular momentum in rotating systems \cite{24,30,31}. Since their introduction by von Zeipel, such surfaces have been extensively studied in the context of rotating stars and black holes, revealing deep connections between geometry, radiation flux, and temperature distributions \cite{30,81,82,83,84,85,86,87,88,89,105,106}.

Motivated by these considerations, the goal of the present work is to investigate the properties of thin accretion disks around a rotating charged black hole described by an effective higher-curvature–inspired spacetime. Adopting a purely kinematical approach, we analyze test-particle motion, ISCO properties, radiative efficiency, flux, temperature, luminosity, frame dragging, and equipotential structures. Our results illustrate how phenomenological deviations from the Kerr–Newman geometry may influence accretion observables, providing a useful framework for interpreting current and future observations of black hole systems.

The structure of the paper is as follows. In Section~\ref{4DEGB}, we introduce the effective rotating charged black hole geometry and discuss its frame-dragging properties. In Section~\ref{TPMaAP}, we analyze test-particle motion and derive the specific energy, angular momentum, and angular velocity relevant to thin accretion disks. The radiative efficiency, flux, temperature, and luminosity of the disk are studied in detail. Section~\ref{RFoP} is devoted to rotational effects in strong gravitational fields and the analysis of equipotential and von Zeipel surfaces. Finally, Section~\ref{SaC} summarizes our results and presents concluding remarks.

\section{Effective Rotating Charged Black Hole Geometry with Gauss-Bonnet-Like Corrections}
\label{4DEGB}

General Relativity, as a classical theory of gravitation, has proven remarkably successful in describing gravitational phenomena across a wide range of scales. Nevertheless, it is widely acknowledged that GR faces conceptual and phenomenological challenges both at very small scales—where a consistent quantum description of gravity is still lacking—and at cosmological scales, where observations such as the late-time accelerated expansion of the Universe and galactic rotation curves are commonly attributed to dark energy and dark matter components. These issues have motivated extensive investigations into gravitational models that incorporate corrections to the Einstein–Hilbert framework.

Among the most studied extensions are theories involving higher-curvature terms, such as Gauss–Bonnet–type corrections, which naturally arise in low-energy limits of string theory and other quantum-gravity-inspired approaches. In four spacetime dimensions, however, the Gauss–Bonnet term is purely topological and does not contribute dynamically to the field equations unless additional degrees of freedom are introduced. While several formulations have been proposed to construct four-dimensional theories with Gauss–Bonnet–like effects, it is now well established that consistent realizations generally involve additional scalar degrees of freedom and that rotating black hole solutions cannot, in general, be generated through the Newman–Janis algorithm.

In view of these facts, we do not attempt to construct or analyze an exact solution of a fundamental four-dimensional Einstein–Gauss–Bonnet theory. Instead, we adopt a phenomenological and kinematical approach, in which the spacetime geometry is treated as an effective rotating charged black hole metric inspired by Gauss–Bonnet–type higher-curvature corrections. Such effective geometries are commonly employed in studies of strong-gravity observables, allowing one to investigate how deviations from the Kerr geometry may influence particle dynamics and astrophysical signatures without committing to a specific underlying gravitational action.

\subsection{ Effective Metric Ansatz}

We consider a stationary and axisymmetric spacetime whose line element is written in Boyer–Lindquist–like coordinates as \cite{47}
\begin{equation}\label{ds2_eff}
\begin{aligned}
ds^2 ={}& -\frac{\Delta}{\rho^2}\left(dt - a \sin^2\theta \, d\phi \right)^2
+ \frac{\rho^2}{\Delta} dr^2
+ \rho^2 d\theta^2  \\
&+ \frac{\sin^2\theta}{\rho^2}
\left( a\, dt - (r^2 + a^2)\, d\phi \right)^2 ,
\end{aligned}
\end{equation}
where the metric functions are defined as
\begin{equation}
\rho^2 = r^2 + a^2 \cos^2\theta ,
\end{equation}
\begin{equation}\label{Delta_eff}
\Delta = r^2 + a^2 + \frac{r^2}{2\alpha}
\left[
1 - \sqrt{1 + 4\alpha \left( \frac{2M}{r^3} - \frac{Q^2}{r^4} \right)}
\right] .
\end{equation}

Here, $M$ denotes the mass parameter, $a$ is the rotation parameter, $Q$ represents an effective charge parameter, and $\alpha$ encodes deviations from the Kerr–Newman geometry associated with higher-curvature–inspired corrections. We emphasize that the parameter $\alpha$ should be interpreted phenomenologically, rather than as a coupling constant of a specific covariant four-dimensional Gauss–Bonnet theory.

The nonvanishing components of the metric tensor are given by
\begin{equation}\label{Metric_eff}
\left\{
\begin{aligned}
g_{tt} &= \frac{1}{\rho^2}\left(-\Delta + a^2 \sin^2\theta\right), \\
g_{t\phi} &= \frac{2a \sin^2\theta}{\rho^2}
\left(\Delta - (r^2 + a^2)\right), \\
g_{\phi\phi} &= \frac{\sin^2\theta}{\rho^2}
\left[ (r^2 + a^2)^2 - a^2 \Delta \sin^2\theta \right], \\
g_{\theta\theta} &= \rho^2, \\
g_{rr} &= \frac{\rho^2}{\Delta}.
\end{aligned}
\right.
\end{equation}

It is important to stress that the metric~(\ref{ds2_eff}) is not required to satisfy the vacuum field equations of any specific four-dimensional higher-curvature gravity theory. In particular, the Newman–Janis algorithm employed in earlier works does not, in general, generate exact rotating solutions of consistent four-dimensional Gauss–Bonnet models, and the resulting geometry may be interpreted as being supported by an effective stress–energy tensor. In the present work, we do not attempt to model this effective matter content explicitly.

Our analysis is therefore purely kinematical, focusing on test-particle motion and thin accretion disk properties in the above effective spacetime. This approach allows us to isolate and quantify the impact of higher-curvature–inspired deformations of the Kerr geometry on observable accretion disk characteristics, such as the innermost stable circular orbit, radiative efficiency, flux, temperature, and luminosity.

The location of the event horizon is determined by the condition
\begin{equation}
\Delta = 0 .
\end{equation}
Depending on the values of the parameters $a$, $Q$, and $\alpha$, the spacetime may admit two horizons (an inner Cauchy horizon and an outer event horizon), a single degenerate horizon, or no horizon, corresponding to a naked singularity. As in the Kerr–Newman case, increasing the rotation or charge parameters generally leads to a reduction in the radius of the outer event horizon. In the following sections, we restrict our analysis to parameter ranges for which a well-defined event horizon exists.

\subsection{Frame Dragging}

Frame-dragging is an effect on space-time caused by non-static (e.g., rotating) fixed mass-energy distributions in a stationary field, predicted by Einstein's General Theory of Relativity. Einstein's 1913 letter to Mach about the relativistic Coriolis force of the Foucault pendulum led to the first use of the term dragging''. Then, Lense and Thirring in their paper introduced the dragging coefficient'' as the (inverse) ratio between the angular velocity of the shell ($\Omega$) and the angular velocity of the reference frame ($\Omega'$) within which the Coriolis effect (or force) becomes zero \cite{50,51,52}. It should be noted that the term ``dragging of inertial frame'' was first used in 1965 by Cohen \cite{48}.

The basic principle for understanding frame-dragging is analogous to the Coriolis force experienced by a rotating object in a non-inertial frame. The Coriolis force is a force in a non-inertial frame that causes the motion of objects to deviate along the rotational coordinate axis relative to an inertial frame. Newton's laws of motion describe the motion of objects in inertial frames; however, when the motion of an object is generalized to rotating coordinates, Coriolis and centrifugal forces appear. The Coriolis force is obtained from the cross product of the angular velocity of a coordinate system and the projection of the particle's velocity onto a plane perpendicular to the axis of rotation \cite{108}. The centrifugal force acts outward in the direction perpendicular to the rotational axis and is proportional to the object's distance from the axis of rotation. These forces are collectively referred to as ``fictitious forces'' (or inertial forces) and effectively enable the application of Newton's laws in rotating reference frames, representing correction terms absent in non-accelerated (inertial) frames \cite{53}.

In essence, frame-dragging describes how the rotation of a massive object twists the surrounding spacetime, causing nearby objects to undergo precessional motion. Related aspects of this phenomenon appear in other areas of physics. For example, in axisymmetric stationary spacetimes, observers with zero angular momentum (ZAMOs) possess a nonzero angular velocity relative to distant inertial observers; this effect is commonly referred to as ``dragging of the ZAMOs'' \cite{49}. Two complementary approaches are often used to interpret frame-dragging effects: an electromagnetic analogy and a fluid-dragging analogy. The electromagnetic analogy is based on the equation of state of solids in the slow-motion and weak-field limits, and it relates the magnetic field to the general relativistic Coriolis (gravitomagnetic) field. The fluid-dragging analogy compares frame-dragging to the rotational effects experienced by an object immersed in a viscous fluid \cite{54}. It should be noted that Rindler, by examining these analogies across different theories and highlighting inconsistencies in the fluid-dragging picture, argued that the gravitomagnetic analogy provides a more reliable interpretation \cite{55}.

The frame-dragging angular velocity for the effective rotating charged spacetime considered in this work is given by \cite{23}:

\begin{equation}\label{frame_dragging_01}
  \omega  =  - \frac{{{\eta ^\nu }{\xi _\nu }}}{{{\xi ^\mu }{\xi _\mu }}} =  - \frac{{{g_{t\phi }}}}{{{g_{\phi \phi }}}} = \frac{2}{{a\left( {{{\sin }^2} \theta   - \frac{{\left( {{a^2} + {r^2}} \right)\left( {{a^2} + {r^2} - {{\sin }^2} \theta  } \right)}}{{\frac{{{r^2}}}{{2\alpha }}\left[ {1 - \sqrt {1 - \frac{{4\left( {{Q^2} - 2Mr} \right)\alpha }}{{{r^4}}}} } \right]}}} \right)}} \,,
\end{equation}
where ${\eta ^\mu } = \delta _t^\mu $ and ${\xi ^\mu } = \delta _\phi ^\mu $ are two commuting (timelike and axial) Killing vectors.

\begin{figure}[htb]
\centering
\subfloat{\includegraphics[width=0.475\textwidth]{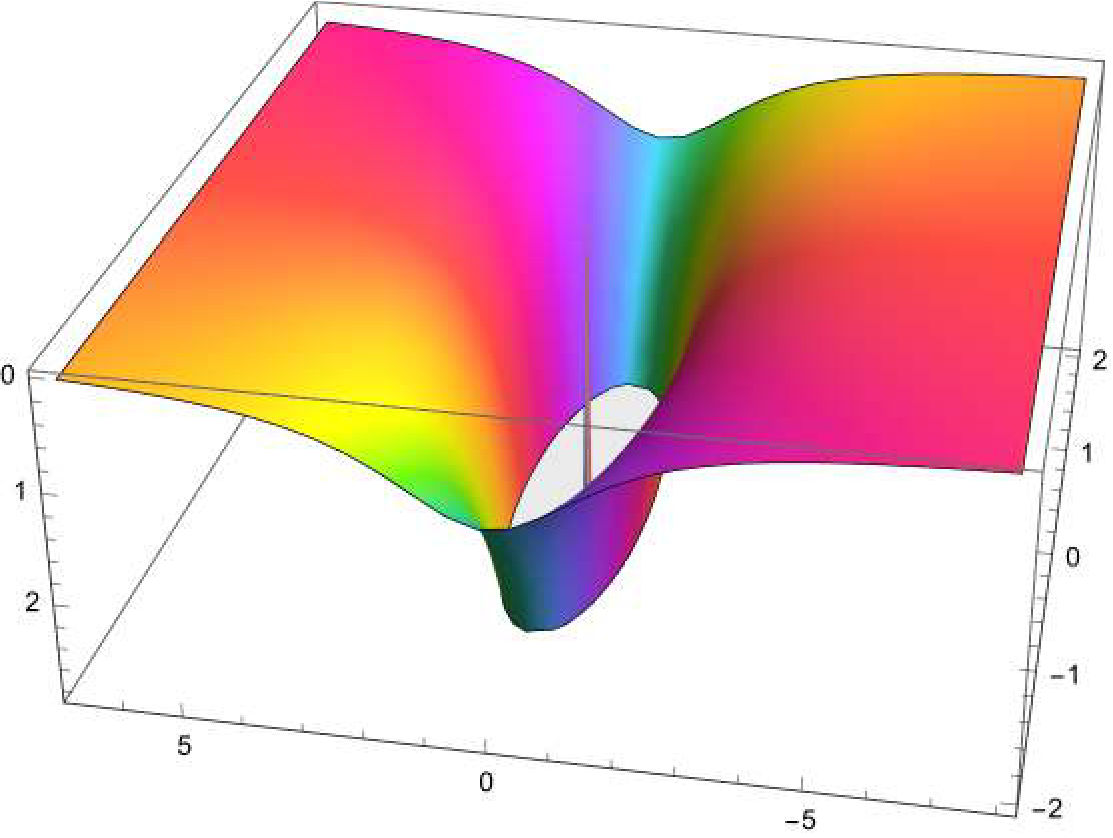}}
\,\,\,
\subfloat{\includegraphics[width=0.475\textwidth]{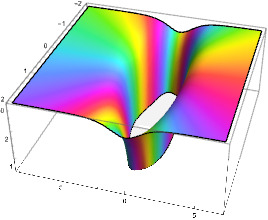}}
\,\,\,
\subfloat{\includegraphics[width=0.475\textwidth]{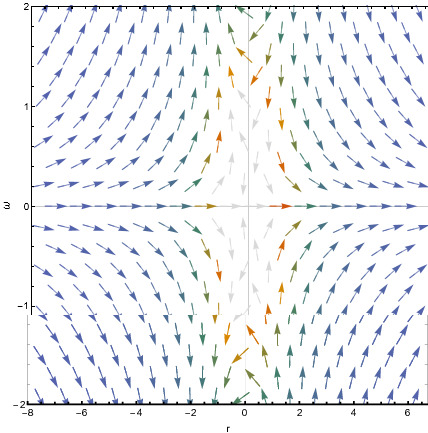}}
\,\,\,
\subfloat{\includegraphics[width=0.475\textwidth]{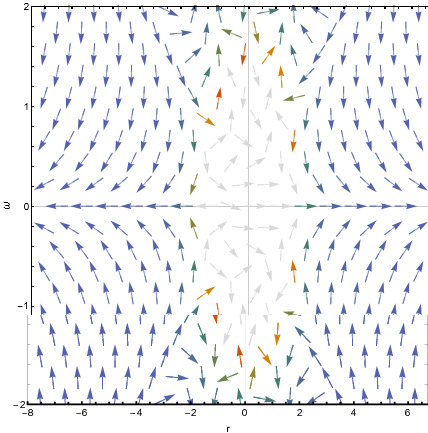}}
\,\,\,
\subfloat[\label{Kerr_Frame_Dragging} Frame dragging for kerr black hole]{\includegraphics[width=0.475\textwidth]{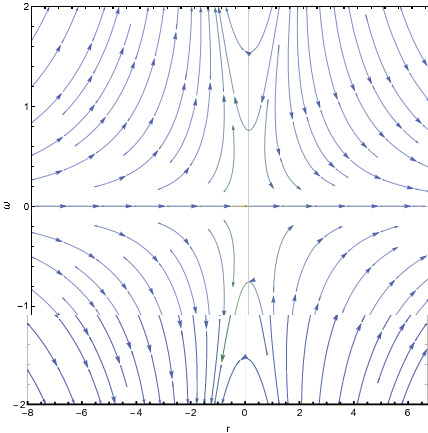}}
\,\,\,
\subfloat[\label{RC_4D_EGB_black_hole} Frame dragging for RC 4D EGB balck hole]{\includegraphics[width=0.475\textwidth]{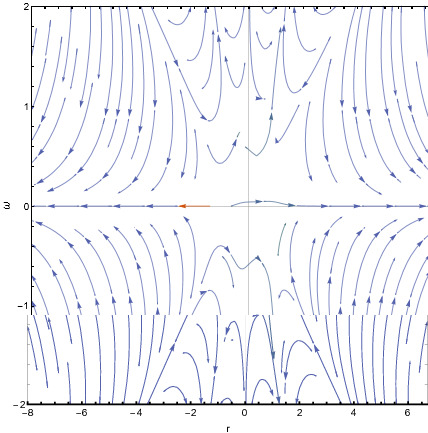}}
\caption{\label{Frme_Dragging}\small{\emph{Plots of the frame-dragging angular velocity for a comparison between the Kerr black hole and the effective rotating charged black hole spacetime considered in this work. The horizontal axis represents the radial distance, while the vertical axis shows how spacetime rotates around the black hole, with positive (negative) values indicating rotation in the same (opposite) direction as the black hole’s spin..}}}
\end{figure}

To enhance our understanding of frame-dragging effects, we plot the frame-dragging angular velocity for the Kerr black hole and for the effective rotating charged spacetime considered in this work in Fig.~\ref{Frme_Dragging}. The figure illustrates the distortion of spacetime, or equivalently of particle geodesics, induced by rotation and charge in the effective geometry. Specifically, the plot shows the behavior of the frame-dragging angular velocity $\omega$ as a function of the radial distance, with the vertical axis representing $\omega$ and the horizontal axis denoting the radial coordinate.

Positive (negative) values of $\omega$ indicate that spacetime rotates in the same (opposite) direction as the black hole’s spin. As the radial distance decreases and one approaches the black hole ($r \to r_H$), the magnitude of $\omega$ increases, reflecting the strengthening of frame-dragging effects in the strong-field region. Near the event horizon, the sign and magnitude of $\omega$ depend on the direction of rotation, and the change in slope of the $\omega$ curve highlights the pronounced influence of rotation on the surrounding spacetime.

Compared to the Kerr case, the effective rotating charged spacetime exhibits a shift of the event horizon toward smaller radii and a modification of the radial gradient of the frame-dragging angular velocity. These features arise from the effective higher-curvature--inspired deformation of the geometry and indicate quantitative differences in rotational behavior in the near-horizon region, without invoking a specific underlying field-theoretic description.

ZAMO observers, which define a natural locally non-rotating reference frame in stationary and axisymmetric spacetimes, follow timelike worldlines with zero angular momentum. Such observers move orthogonally to hypersurfaces of constant time under the influence of gravitational acceleration. The four-velocity of ZAMO observers is defined as follows:

\begin{equation*}
  {N^\mu } \equiv {e^{ - \Phi }}{{\tilde \eta }^\mu }\,.
\end{equation*}
where ${{\tilde \eta }^\mu } = {\eta ^\nu } + \omega {\xi ^\mu }$ (the modified Killing vector to remove background rotation). Since we know that ZAMO’s observers formulates the reference frame in 3-dimensional space, we can write the gravitational potential ($\Phi$) in the ZAMO’s frame as follows \cite{23}:

\begin{equation}\label{gravitational_potential_01}
  \Phi  =  - \frac{1}{2}\ln \left[ {\frac{{{\xi ^\mu }{\xi _\mu }}}{{{{\left( {{\eta ^\nu }{\xi _\nu }} \right)}^2} - \left( {{\eta ^\mu }{\eta _\mu }} \right)\left( {{\xi ^\mu }{\xi _\mu }} \right)}}} \right] =  - \frac{1}{2}\ln \left[ {{g^{tt}}} \right]\,.
\end{equation}
In Newton's theory, the force applied by gravity's acceleration is equal to the gradient of the gravitational potential. However, it is seen that in the surface $\frac{1}{{{g^{tt}}}} = 0$, ${{\tilde \eta }^\mu }$ is null vector; that is, it moves at the speed of light on the surface.

The effective curvature parameter $\alpha$ introduced in this work is treated as a phenomenological quantity encoding higher-curvature corrections to the Kerr--Newman geometry. In the absence of a fully consistent four-dimensional Einstein--Gauss--Bonnet framework, $\alpha$ should not be interpreted as a fundamental coupling constant, but rather as an effective parameter characterizing deviations from General Relativity. Theoretical consistency requires that $\alpha$ remains within a range that preserves the regularity of the metric function and ensures the existence of a well-defined event horizon. In particular, the condition that the argument of the square root in the metric function remains positive imposes restrictions on admissible values of $\alpha$.

From an observational perspective, current measurements of black hole environments, including shadow observations and X-ray spectroscopy, indicate that deviations from the Kerr geometry are tightly constrained. This suggests that $\alpha$ should be relatively small, corresponding to perturbative corrections to the standard spacetime. In this work, we therefore consider representative values of $\alpha$ that allow us to explore qualitative trends in accretion disk properties, while a detailed derivation of quantitative observational bounds is deferred to future studies.

\section{Test Particle Motion and Accretion Properties}\label{TPMaAP}

Since the metric components just depend on $r$ and $\theta$, this metric has two Killing vectors ${\partial _t}$ and ${\partial _\phi }$. Therefore, we can define two constants of motion, energy and angular momentum per unit mass (specific energy and specific angular momentum, respectively) as follows \cite{58}:

\begin{equation}\label{E&L_tilda1}
    \frac{d}{{d\tau }}\left( {\frac{{\partial {\cal L}}}{{\partial{{\dot x}^\mu }}}} \right) - \frac{{\partial {\cal L}}}{{\partial{x^\mu }}} = 0 \Rightarrow \left\{ {\begin{array}{*{20}{c}}
    {{p_t} \equiv \frac{{\partial {\cal L}}}{{\partial\dot t}} = \tilde E =  - \left[ {{g_{tt}}\dot t + {g_{t\phi }}\dot \phi } \right]}\\
    {}\\
    {{p_\phi } \equiv \frac{{\partial {\cal L}}}{{\partial\dot \phi }} = \tilde L = \left[ {{g_{t\phi }}\dot t + {g_{\phi \phi }}\dot \phi } \right]}
    \end{array}} \right.\,.
\end{equation}

Now we can obtain the 4-vector components of the test particle's velocity as follows:

\begin{equation}\label{4_velocitycomponent_01}
\left\{ {\begin{array}{*{20}{c}}
{\dot t = \frac{{\tilde E{g_{\phi \phi }} + \tilde L{g_{t\phi }}}}{{g_{t\phi }^2 - {g_{tt}}{g_{\phi \phi }}}}}\\
{}\\
{\dot \phi  = \frac{{\tilde E{g_{t\phi }} + \tilde L{g_{tt}}}}{{g_{t\phi }^2 - {g_{tt}}{g_{\phi \phi }}}}}
\end{array}} \right.
\end{equation}
where, $\tilde E$ and $\tilde L$ are the conserved energy and angular momentum per unit mass, respectively. The 4-velocity component equations are fundamental to particle dynamics in static, axisymmetric spacetimes and these equations are the basis to analyze the accretion flows in strong gravitational fields.

\subsection{Descriptive models of the accretion disk}

To understand the mechanism of black hole accretion, we need to explain this phenomenon based on existing reliable models. For this purpose, one can use two major models. In the first model, one uses the energy distribution of particles in the accretion process, and in the second model, one considers the equation of state and density of a fluid (ideal fluid, perfect fluid, dust, etc.) to obtain the radial velocity and energy density in the accretion process. We proceed with the first model assuming the Novikov-Thorne criteria (the accretion disk is thin, optically thick, and modeled in a relativistic medium) and the second model assuming circular accretion of particles as Bondi accretion (particles move inward on the spherical accretion disk) \cite{98}.

For this purpose, we consider the accretion disk as a thin, optically thick circular plate in the equatorial plane vertical to the black hole rotation axis and in a relativistic medium. Therefore, the geodesics of particle motion in the equatorial circular orbits are as follows:

\begin{equation}\label{geodesic_01}
  \frac{d}{{d\tau }}\left( {{g_{\mu \nu }}{{\dot x}^\nu }} \right) = \frac{1}{2}\left( {{\partial _\mu }{g_{\nu \rho }}} \right){{\dot x}^\nu }{{\dot x}^\rho }\,.
\end{equation}

By assuming $\dot r = \dot \theta  = \ddot r = 0$, we can obtain the radial component of polar circular orbits as follows:

\begin{equation}\label{radial_component_01}
  \left( {{\partial _r}{g_{tt}}} \right){{\dot t}^2} + 2\left( {{\partial _r}{g_{t\phi }}} \right)\dot t\dot \phi  + \left( {{\partial _r}{g_{\phi \phi }}} \right){{\dot \phi }^2} = 0\xrightarrow{{\Omega  = \frac{{\dot t}}{{\dot \phi }}}}\left( {{\partial _r}{g_{\phi \phi }}} \right){\Omega ^2} + 2\left( {{\partial _r}{g_{t\phi }}} \right)\Omega  + \left( {{\partial _r}{g_{tt}}} \right) = 0\,.
\end{equation}

Since the effective rotating charged spacetime considered in this work is written in a form analogous to Boyer–Lindquist coordinates, \cite{47}, we need the following relations to calculate the specific energy ($\tilde E$), specific angular momentum ($\tilde L$), and angular velocity ($\Omega$) in the equatorial plane ($\theta  = \frac{\pi }{2} \leftrightarrow \sin \theta  = 1 , \cos \theta  = 0$):
\begin{equation}\label{Delta_01}
  \left\{ {\begin{array}{*{20}{c}}
  {\Delta  = {r^2} + {a^2} + \frac{{{r^2}}}{{2\alpha }}\left[ {1 - \sqrt {1 + 4\alpha \left( {\frac{{2M}}{{{r^3}}} - \frac{{{Q^2}}}{{{r^4}}}} \right)} } \right]}\,, \\
  {}\\
  {{\rho ^2} = {r^2} + {a^2}{{\cos }^2}\theta \xrightarrow{{\theta  = \frac{\pi }{2}}}{\rho ^2} = {r^2}}\,.
\end{array}} \right.
\end{equation}

The metric components are simplified as follows.
\begin{equation}\label{metric_01}
  \left\{ {\begin{array}{*{20}{c}}
  {{g_{tt}} = \frac{{{a^2} - \Delta }}{{{r^2}}}}\,, \\
  {}\\
  {{g_{t\phi }} = \frac{{2a}}{{{r^2}}}\left( {\Delta  - \left( {{r^2} + {a^2}} \right)} \right)}\,, \\
  {}\\
  {{g_{\phi \phi }} = \frac{{{a^2}}}{{{r^2}}}\left( {{{\left( {{r^2} + {a^2}} \right)}^2} - \Delta } \right)}\,.
\end{array}} \right.
\end{equation}

So, the derivative of the metric components are obtained as follows:
\begin{equation}\label{Delta_02}
  \left. {\begin{array}{*{20}{c}}
  {A = 1 + 4\alpha \left( {\frac{{2M}}{{{r^3}}} - \frac{{{Q^2}}}{{{r^4}}}} \right)} \\
  {}\\
  {{\partial _r}A = 4\alpha \left( {\frac{{6M}}{{{r^4}}} - \frac{{4{Q^2}}}{{{r^5}}}} \right)}
\end{array}} \right\}\Rightarrow{\partial _r}\Delta  = 2r + \frac{1}{{2\alpha }}\left[ {2r\left( {1 - \sqrt A } \right) + {r^2}\left( { - \frac{1}{{2\sqrt A }}{\partial _r}A} \right)} \right]\,,
\end{equation}
where
\begin{equation}\label{metric_02}
  \left\{ {\begin{array}{*{20}{c}}
  {{\partial _r}{g_{tt}} = \frac{{ - r\left( {{\partial _r}\Delta } \right) - 2\left( {{a^2} - \Delta } \right)}}{{{r^3}}}}\,, \\
  {}\\
  {{\partial _r}{g_{t\phi }} = \frac{{2a}}{{{r^3}}}\left( {r\left( {{\partial _r}\Delta } \right) - 2\Delta  + 2{a^2}} \right)}\,, \\
  {}\\
  {{\partial _r}{g_{\phi \phi }} = \frac{{{a^2}}}{{{r^3}}}\left( {2{r^4} + 4{a^2}{r^2} + 2{a^4} - {r^2}\left( {{\partial _r}\Delta } \right) + 2r\Delta } \right)}\,.
\end{array}} \right.
\end{equation}

Now, we can obtain the specific energy, specific angular momentum, and angular velocity using the assumptions $\dot r = \dot \theta  = \ddot r = 0$ as well as the velocity normalization condition ${{\dot x}^\mu }{{\dot x}_\mu } =  - 1$ \cite{58,59}:

\begin{equation}\label{angular_velocity_01}
  {{\rm{\Omega }}_ \pm } = \frac{{ - {\partial _r}{g_{t\phi }} \pm \sqrt {{{\left( {{\partial _r}{g_{t\phi }}} \right)}^2} - \left( {{\partial _r}{g_{tt}}} \right)\left( {{\partial _r}{g_{\phi \phi }}} \right)} }}{{{\partial _r}{g_{\phi \phi }}}}
\end{equation}
where using equations \eqref{metric_02}, we find:
\begin{equation}\label{angular_velocity_03}
  {{\rm{\Omega }}_ \pm } = \frac{1}{{2a \pm \frac{{{r^3}\sqrt {\frac{{{a^2}\left( {2{a^2} - 2\Delta  + r{\partial _r}\Delta } \right)\left( {2\left( {{a^4} + {r^4} + 2{a^2}\left( {2 + {r^2}} \right) + r\Delta  - 4\Delta } \right) - \left( { - 4 + r} \right)r{\partial _r}\Delta } \right)}}{{{r^6}}}} }}{{2{a^2} - 2\Delta  + r{\partial _r}\Delta }}}}\,.
\end{equation}

\begin{figure}[htb]
\centering
\subfloat[\label{Omega1} The specific angular velocity of particles for the Kerr black hole and the effective rotating and charged rotating spacetimes work.]{\includegraphics[width=0.675\textwidth]{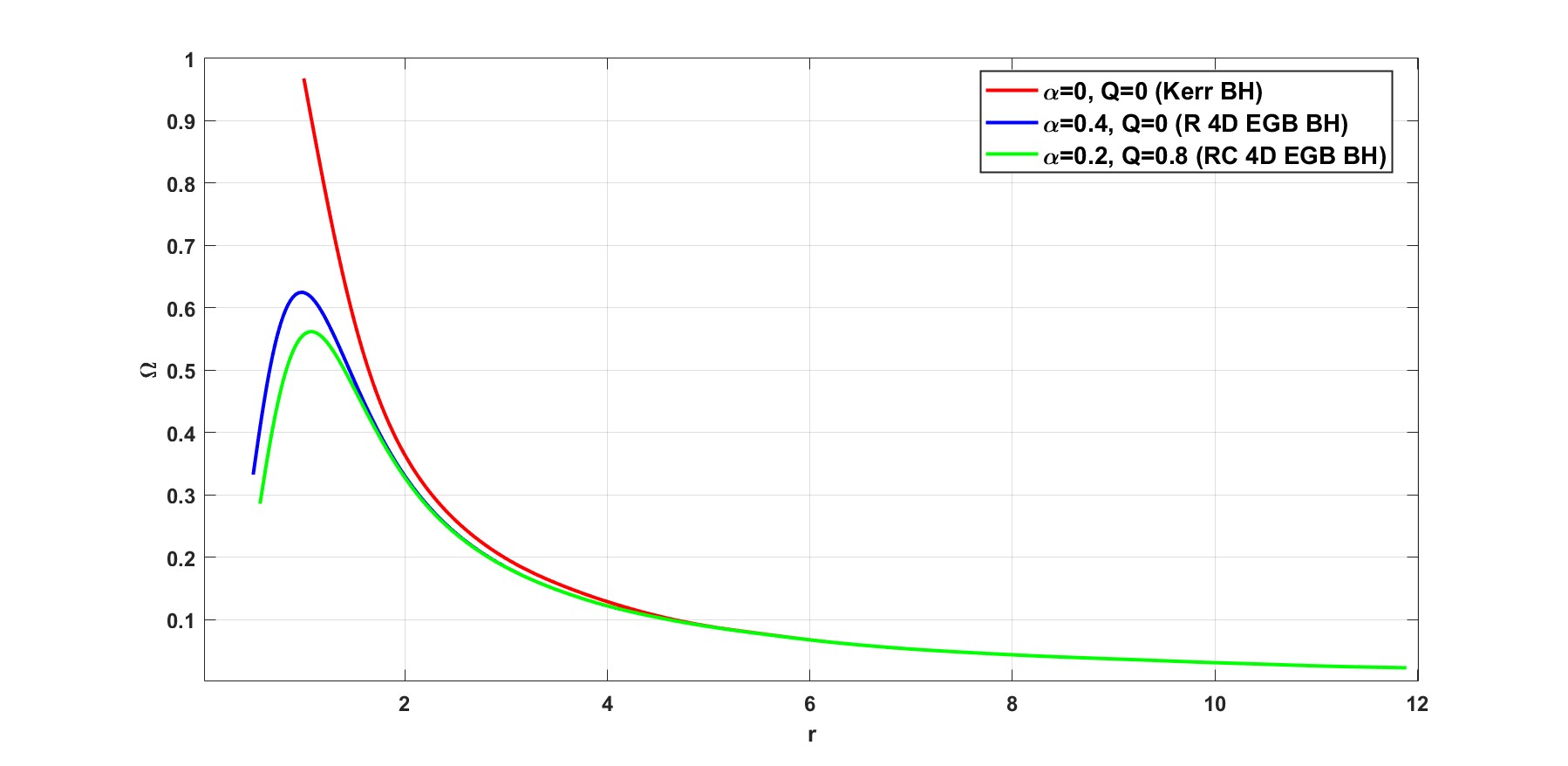}}
\,\,\,
\subfloat[\label{Omega2} Specific angular velocity of particles for different values of $\alpha$ and $Q$ in the effective rotating charged spacetime.
]{\includegraphics[width=0.675\textwidth]{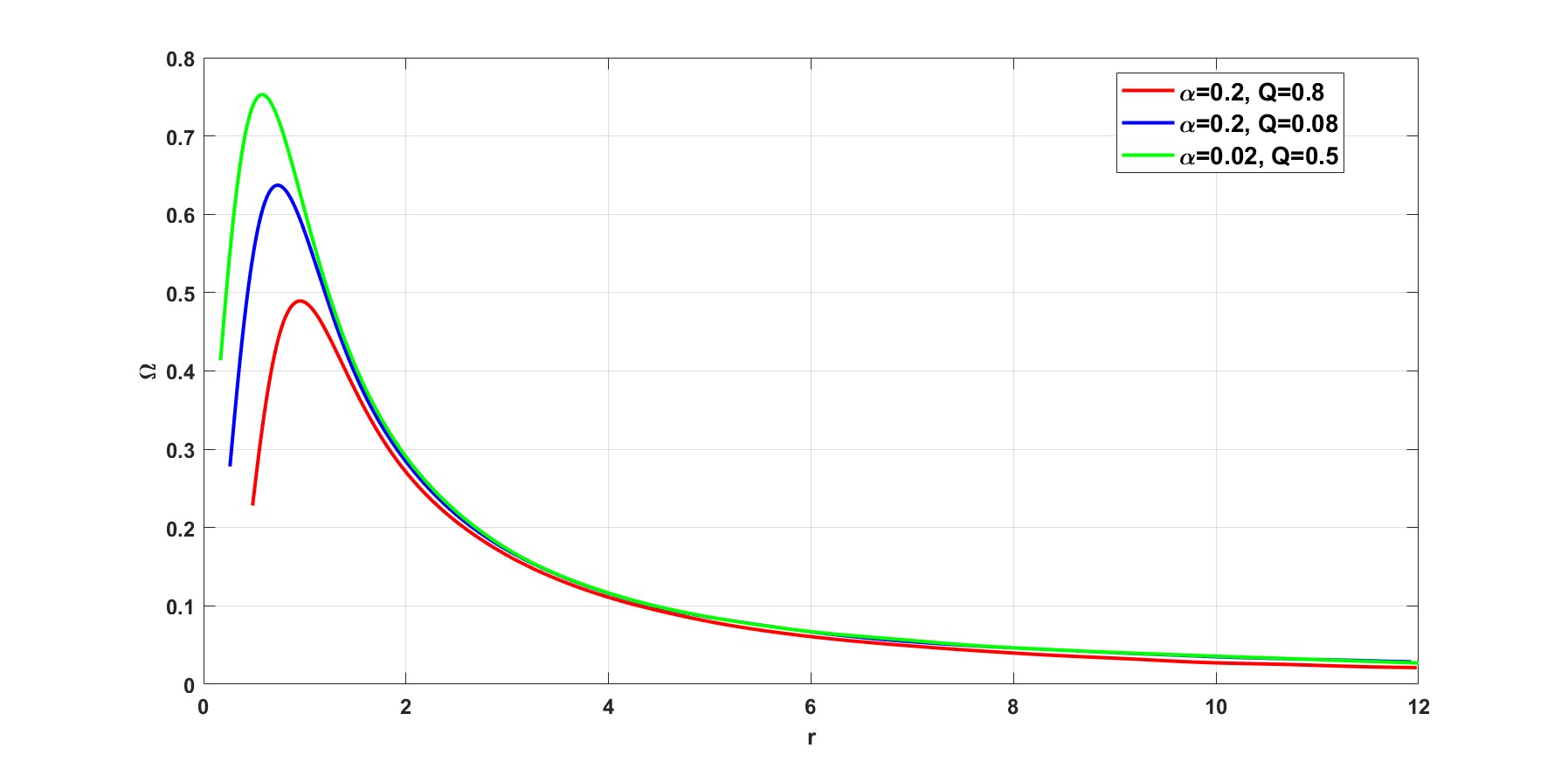}}
\caption{\label{angular_velocity_02}\small{\emph{Figure (Specific angular velocity of particles for the Kerr black hole and the effective rotating charged spacetime considered in this work. The dependence on the effective curvature parameter $\alpha$ and the charge parameter $Q$ is shown for a fixed spin $a = 0.9$.
.}}}
\end{figure}

It should be stressed that ${{\rm{\Omega }}_ + }$ is for particles moving in orbits parallel to the axis of rotation of the black hole and ${{\rm{\Omega }}_ - }$ is for particles moving in orbits opposite to the rotation of the black hole.

For $\tilde E$, we have the following \cite{58,59}:

\begin{equation}\label{specific_energy_01}
  \tilde E =  - \left( {{g_{tt}} + {\rm{\Omega }}{g_{t\phi }}} \right)\dot t =  - \frac{{{g_{tt}} + {\rm{\Omega }}{g_{t\phi }}}}{{\sqrt { - {g_{tt}} - 2{\rm{\Omega }}{g_{t\phi }} - {{\rm{\Omega }}^2}{g_{\phi \phi }}} }}
\end{equation}
where using equations \eqref{metric_02}, we find:
\begin{equation}\label{specific_energy_02}
  \tilde E = \frac{{\frac{{{a^2} - \Delta }}{{{r^2}}} + \frac{{2a\Omega \left( { - {a^2} - {r^2} + \Delta } \right)}}{{{r^2}}}}}{{\sqrt { - \frac{{{a^2} - \Delta }}{{{r^2}}} - \frac{{{a^2}{\Omega ^2}\left( {{{\left( {{a^2} + {r^2}} \right)}^2} - \Delta } \right)}}{{{r^2}}} - \frac{{4a\Omega \left( { - {a^2} - {r^2} + \Delta } \right)}}{{{r^2}}}} }} = \frac{{a\left( {a - 2\left( {{a^2} + {r^2}} \right)\Omega } \right) + \left( { - 1 + 2a\Omega } \right)\Delta }}{{{r^2}\sqrt {\frac{{a\left( { - a + 4\left( {{a^2} + {r^2}} \right)\Omega  - a{{\left( {{a^2} + {r^2}} \right)}^2}{\Omega ^2}} \right) + \left( {1 + a\Omega \left( { - 4 + a\Omega } \right)} \right)\Delta }}{{{r^2}}}} }}
\end{equation}

\begin{figure}[htb]
\centering
\subfloat[\label{E1} Specific energy of particles in the Kerr spacetime and the effective rotating and rotating charged geometries.]{\includegraphics[width=0.675\textwidth]{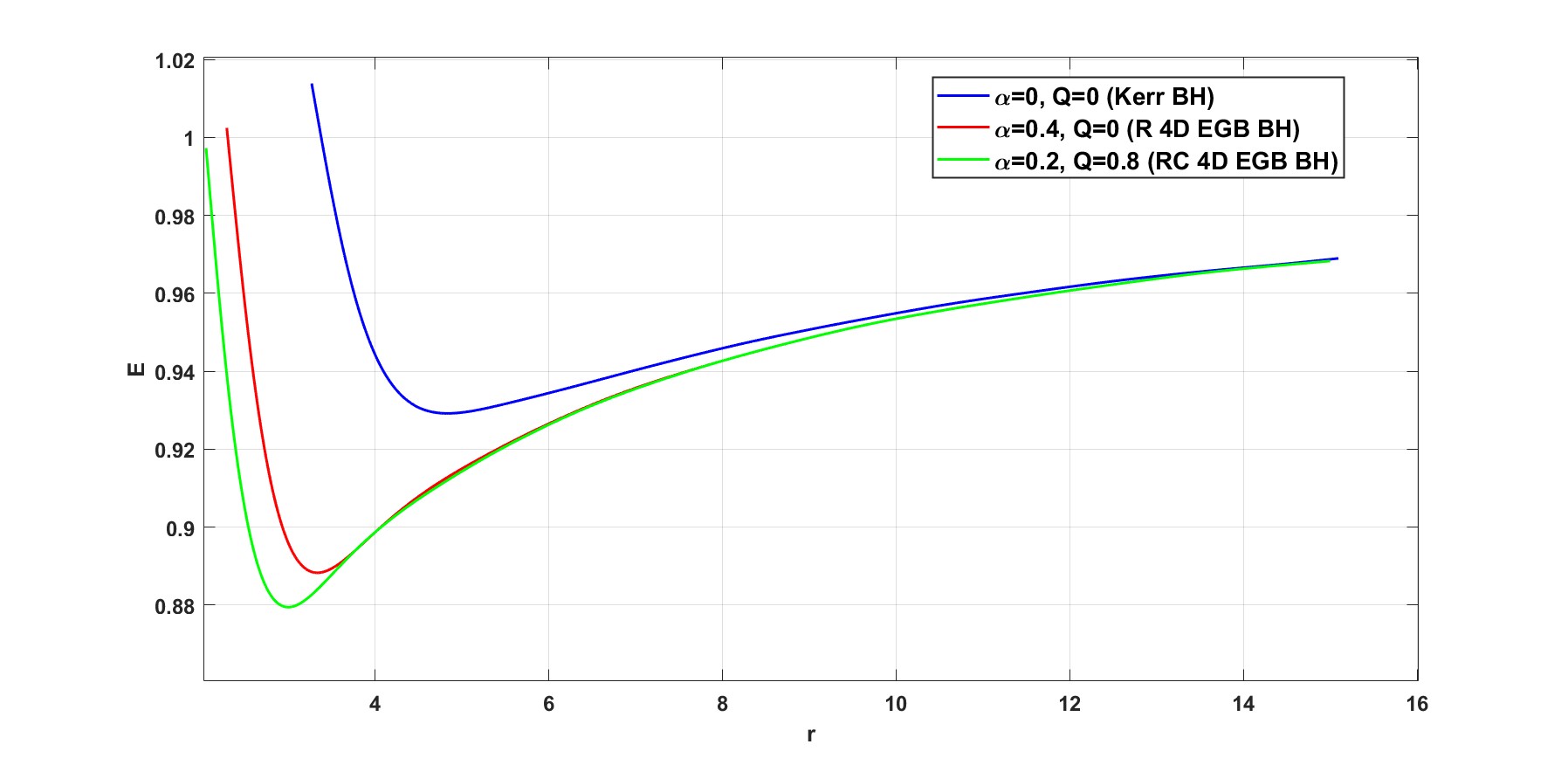}}
\,\,\,
\subfloat[\label{E2} Specific energy of particles for different values of $\alpha$ and $Q$ in the effective rotating charged spacetime.
]{\includegraphics[width=0.675\textwidth]{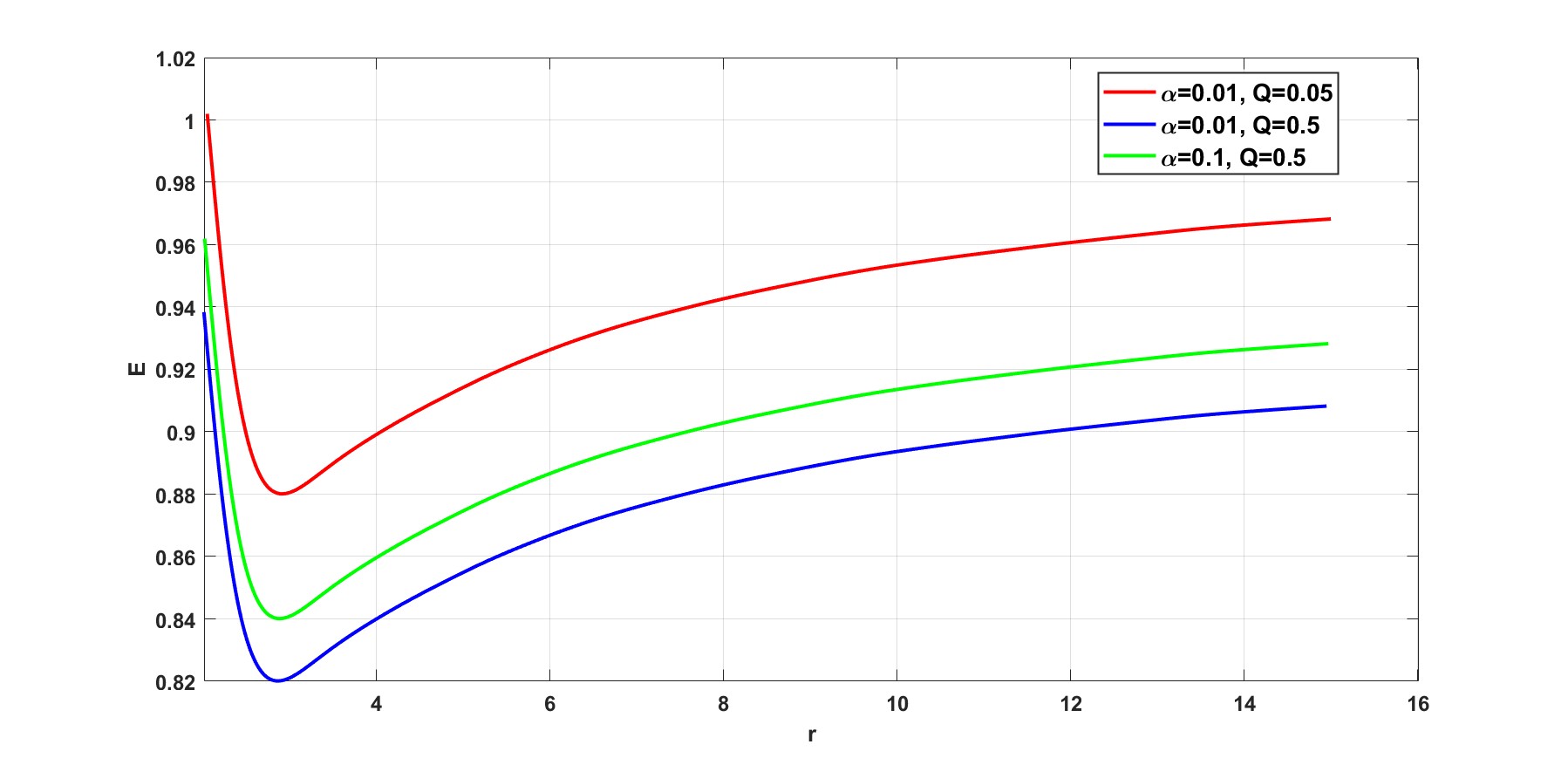}}
\caption{\label{specific_energy_03}\small{\emph{Specific energy of particles for the Kerr black hole and the effective rotating charged spacetime, shown for different values of $\alpha$ and $Q$ with $a = 0.9$.
. }}}
\end{figure}

In similar way, $\tilde L$ can be obtained as follow \cite{58,59}:
\begin{equation}\label{specific_angular_momentum_01}
    \tilde L = \left( {{g_{t\phi }} + {\rm{\Omega }}{g_{\phi \phi }}} \right)\dot t = \frac{{{g_{t\phi }} + {\rm{\Omega }}{g_{\phi \phi }}}}{{\sqrt { - {g_{tt}} - 2{\rm{\Omega }}{g_{t\phi }} - {{\rm{\Omega }}^2}{g_{\phi \phi }}} }}\,,
\end{equation}
where using equations \eqref{metric_02}, we find:
\begin{equation}\label{specific_angular_momentum_02}
  \tilde L = \frac{{\frac{{{a^2}\Omega \left( {{{\left( {{a^2} + {r^2}} \right)}^2} - \Delta } \right)}}{{{r^2}}} + \frac{{2a\left( { - {a^2} - {r^2} + \Delta } \right)}}{{{r^2}}}}}{{\sqrt { - \frac{{{a^2} - \Delta }}{{{r^2}}} - \frac{{{a^2}{\Omega ^2}\left( {{{\left( {{a^2} + {r^2}} \right)}^2} - \Delta } \right)}}{{{r^2}}} - \frac{{4a\Omega \left( { - {a^2} - {r^2} + \Delta } \right)}}{{{r^2}}}} }} = \frac{{a\left( {{a^2} + {r^2}} \right)\left( { - 2 + a\left( {{a^2} + {r^2}} \right)\Omega } \right) + a\left( {2 - a\Omega } \right)\Delta }}{{{r^2}\sqrt {\frac{{a\left( { - a + 4\left( {{a^2} + {r^2}} \right)\Omega  - a{{\left( {{a^2} + {r^2}} \right)}^2}{\Omega ^2}} \right) + \left( {1 + a\Omega \left( { - 4 + a\Omega } \right)} \right)\Delta }}{{{r^2}}}} }}
\end{equation}

\begin{figure}[htb]
\centering
\subfloat[\label{L1} Specific angular momentum of particles in the Kerr spacetime and the effective rotating and rotating charged geometries.]{\includegraphics[width=0.775\textwidth]{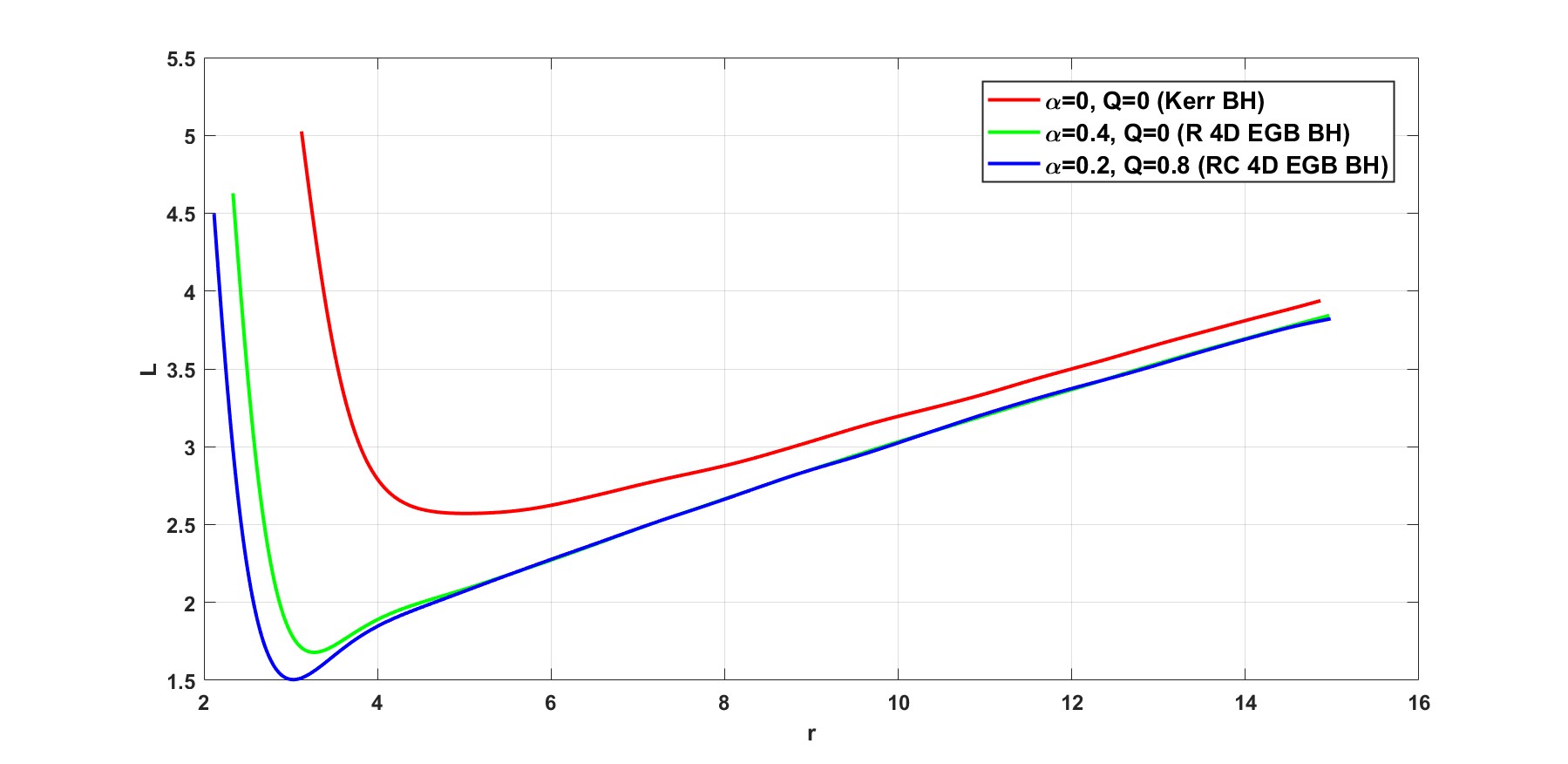}}
\,\,\,
\subfloat[\label{L2} Specific angular momentum of particles for different values of $\alpha$ and $Q$ in the effective rotating charged spacetime.
]{\includegraphics[width=0.775\textwidth]{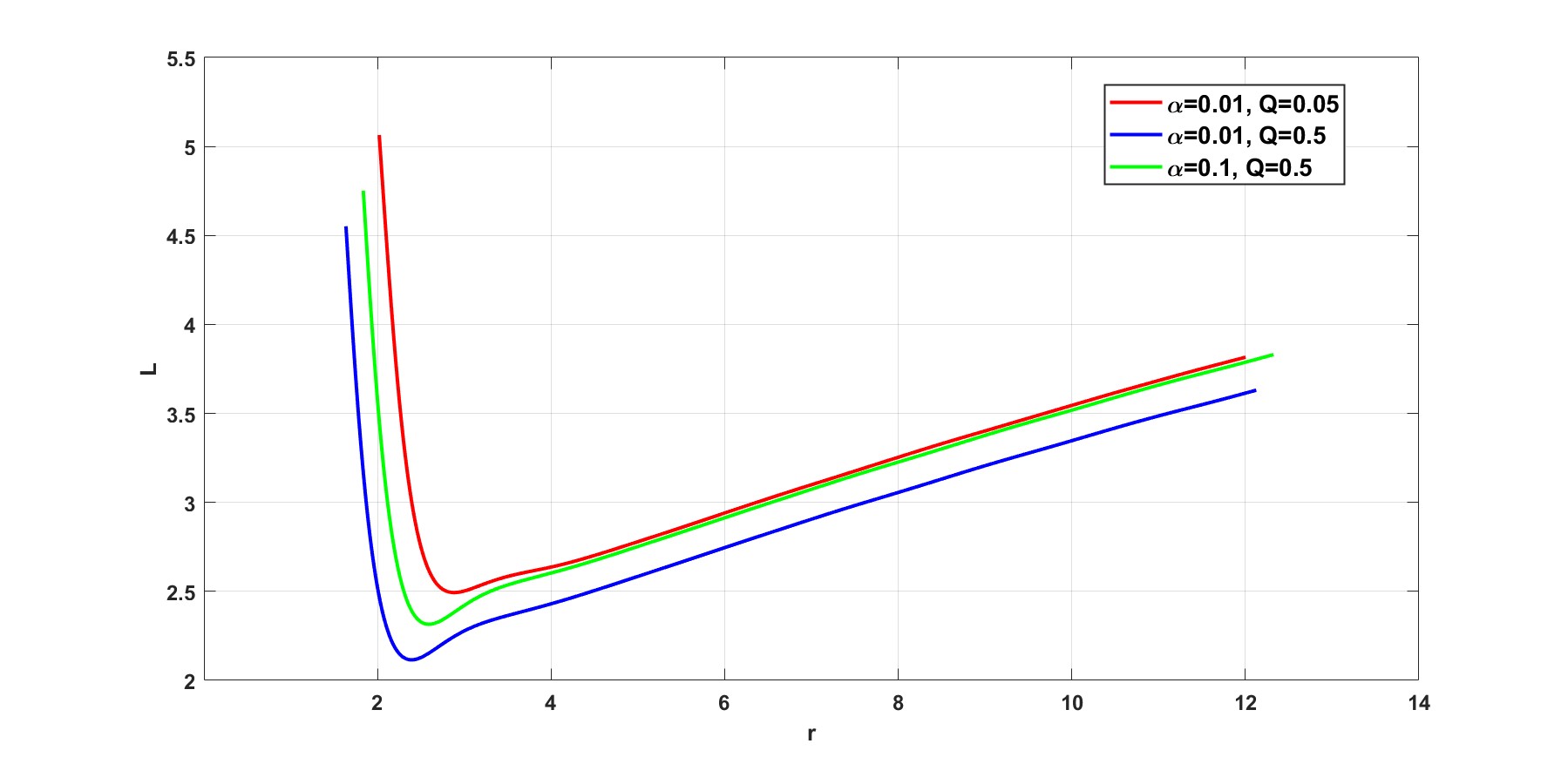}}
\caption{\label{specific_angular_momentum_03}\small{\emph{Specific angular momentum of particles for the Kerr black hole and the effective rotating charged spacetime, shown for different values of $\alpha$ and $Q$ with $a = 0.9$.
.}}}
\end{figure}

The temporal component of the 4-velocity is obtained as follow:
\begin{equation}\label{v_t_01}
  {v^t} = \frac{1}{{\sqrt { - {g_{tt}} - 2{\rm{\Omega }}{g_{t\phi }} - {{\rm{\Omega }}^2}{g_{\phi \phi }}} }}\,,
\end{equation}
where using equations \eqref{metric_02}, we find:
\begin{equation}\label{v_t_02}
  {v^t} = \frac{1}{{\sqrt { - \frac{{{a^2} - \Delta }}{{{r^2}}} - \frac{{4a\left( { - {a^2} - {r^2} + \Delta } \right)\Omega }}{{{r^2}}} - \frac{{{a^2}\left( {{{\left( {{a^2} + {r^2}} \right)}^2} - \Delta } \right){\Omega ^2}}}{{{r^2}}}} }} = \frac{1}{{\sqrt {\frac{{ - {a^2} + \Delta  + a\Delta \Omega \left( { - 4 + a\Omega } \right) + a\left( {{a^2} + {r^2}} \right)\Omega \left( {4 - a\left( {{a^2} + {r^2}} \right)\Omega } \right)}}{{{r^2}}}} }}\,.
\end{equation}

Figure~\eqref{angular_velocity_02} shows the specific angular velocity of particles for the Kerr black hole and for the effective rotating charged spacetime considered in this work. Compared to the Kerr case, the effective geometry generally exhibits a lower specific angular velocity at a given radius, reflecting modifications of the spacetime structure in the strong-field region. Near the event horizon, changes in the geometry alter the angular velocity required for particles to maintain stable circular orbits, particularly in the vicinity of the innermost stable circular orbit (ISCO). The presence of a positive charge parameter introduces an effective electrostatic repulsion that partially counteracts gravitational attraction, leading to a further reduction in the angular velocity required for particles to remain on stable orbits.

Figure~\eqref{E1} illustrates the specific energy of particles for the Kerr black hole and the effective rotating charged spacetime. The specific energy reaches its maximum values close to the event horizon, where gravitational effects are strongest. Compared to the Kerr case, the effective rotating charged spacetime exhibits a lower specific energy for particles on circular orbits. As shown in Fig.~\eqref{E2}, increasing the effective curvature parameter and the charge parameter leads to a further decrease in the specific energy, indicating enhanced energy loss during the accretion process due to geometric deformation and electrostatic effects.

Figure~\eqref{L1} presents the specific angular momentum of particles for the Kerr black hole and the effective rotating charged spacetime. In comparison with the Kerr geometry, the effective spacetime yields a lower specific angular momentum, which is associated with a shift of the ISCO toward smaller radii in the strong-field region. As illustrated in Fig.~\eqref{L2}, increasing the charge parameter results in an additional decrease in the specific angular momentum, reflecting the influence of electrostatic repulsion on particle dynamics.

To obtain the effective potential as an important ingredient to analyze the problem, by applying the above relations to the equation \eqref{E&L_tilda1}, we get

\begin{equation}\label{U_eff_01}
  {U_{eff}}\left( {r,\theta ,\tilde E,\tilde L} \right) = {g_{rr}}{{\dot r}^2} + g_{\theta \theta }^2{{\dot \theta }^2}\,.
\end{equation}

The effective potential functional form is obtained as follows:

\begin{equation}\label{U_eff_02}
  {U_{eff}} = \frac{{{{\tilde E}^2}{g_{\phi \phi }} + 2\tilde E\tilde L{g_{t\phi }} + {{\tilde L}^2}{g_{tt}}}}{{g_{t\phi }^2 - {g_{tt}}{g_{\phi \phi }}}} - 1\,,
\end{equation}
where using equations \eqref{metric_02}, we find:
\begin{eqnarray}\label{U_eff_04}
  {U_{eff}} &=&\frac{{8\left( {{a^2} + {r^2} - \Delta } \right)}}{{\left[ {a\left( {a - 4\left( {{a^2} + {r^2}} \right)\Omega  + a{{\left( {{a^2} + {r^2}} \right)}^2}{\Omega ^2}} \right) - \left( {1 + a\Omega \left( { - 4 + a\Omega } \right)} \right)\Delta } \right]}} \times \nonumber \\
  && \frac{{\left[ {a\left( { - a + 2\left( {{a^2} + {r^2}} \right)\Omega } \right) + \left( {1 - 2a\Omega } \right)\Delta } \right]\left[ {\left( {{a^2} + {r^2}} \right)\left( { - 2 + a\left( {{a^2} + {r^2}} \right)\Omega } \right) + \left( {2 - a\Omega } \right)\Delta } \right]}}{{\left[ {\left( { - 4 + {a^2}} \right){{\left( {{a^2} + {r^2}} \right)}^2} - \Delta \left[ {{a^2}\left( { - 7 + {a^2}} \right) + 2\left( { - 4 + {a^2}} \right){r^2} + {r^4} + 3\Delta } \right]} \right]}} \,.
\end{eqnarray}

\begin{figure}[htb]
\centering
\subfloat[\label{Ueef1} Effective potential in the Kerr spacetime and the effective rotating and rotating charged geometries.]{\includegraphics[width=0.775\textwidth]{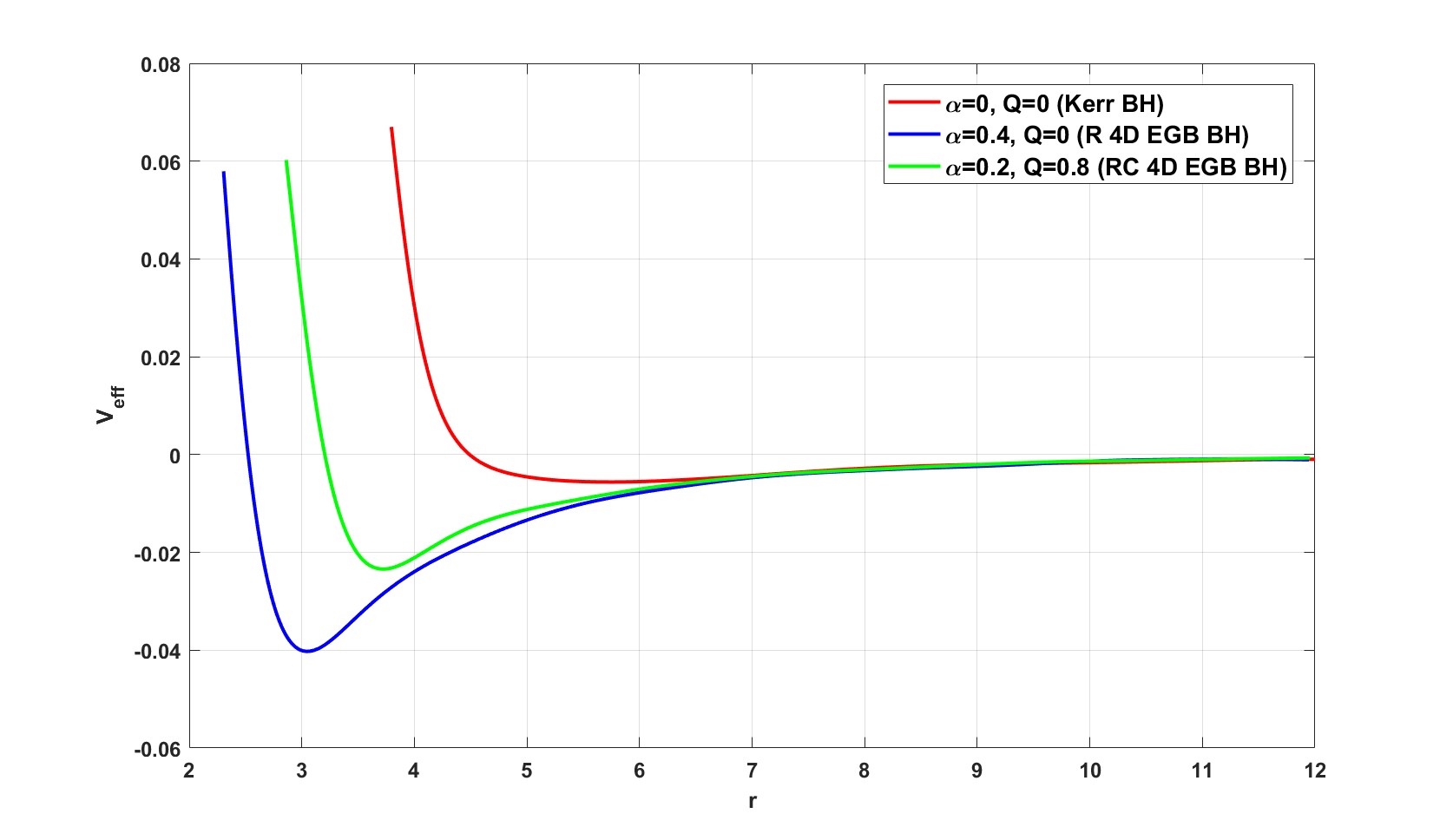}}
\,\,\,
\subfloat[\label{Ueef2} Effective potential for different values of $\alpha$ and $Q$ in the effective rotating charged spacetime.
]{\includegraphics[width=0.775\textwidth]{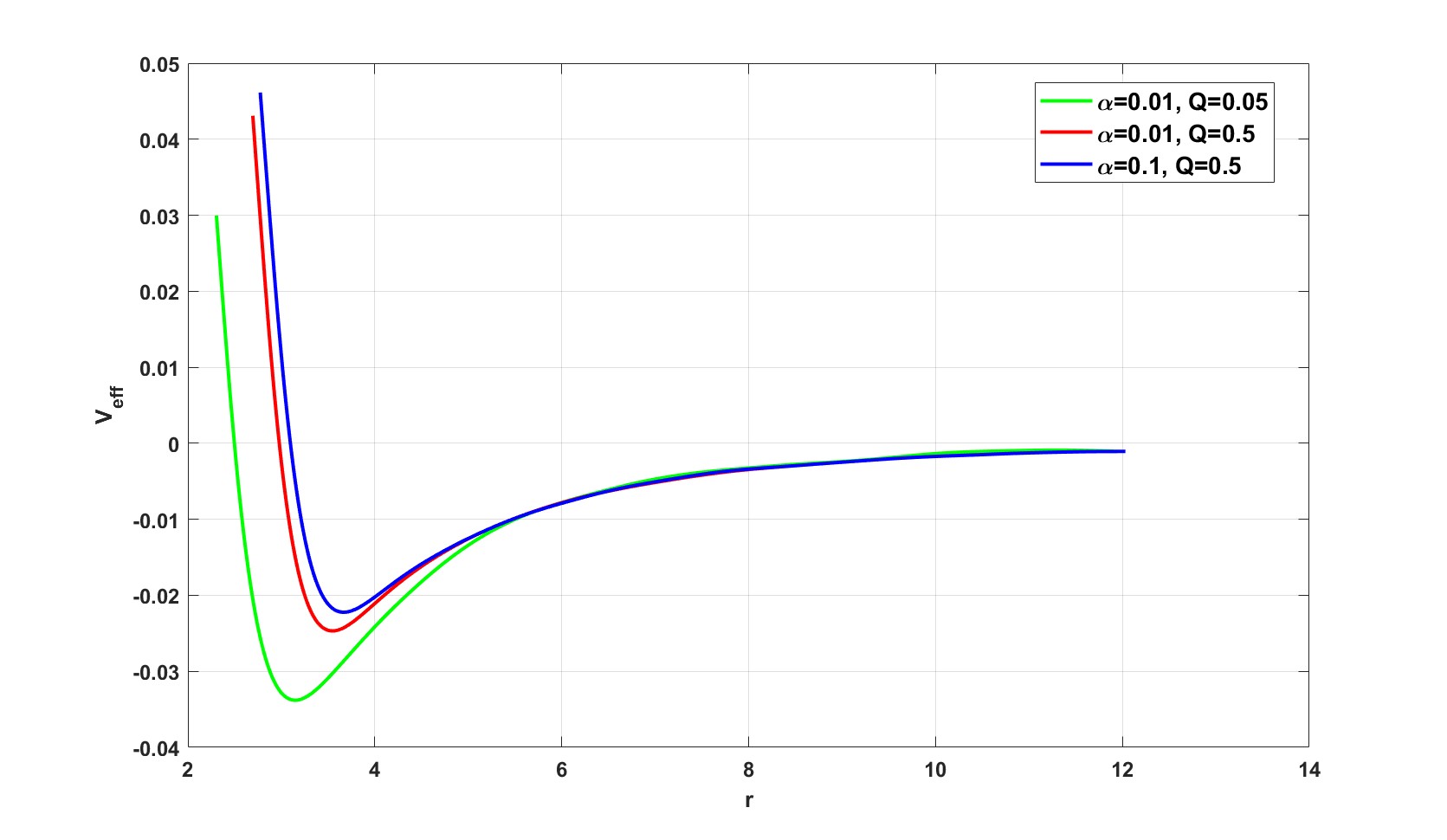}}
\caption{\label{U_eff_03}\small{\emph{Effective potential of particles for the Kerr black hole and the effective rotating charged spacetime, shown for different values of $\alpha$ and $Q$ with $a = 0.9$.
}}}
\end{figure}

Since we want to find the circular orbits, we must satisfied the conditions $\dot r = \ddot r = \dot \theta  = \ddot \theta  = 0$. The radial location that the effective potential’s local is minimum is known as  innermost (marginally) stable circular orbit (ISCO) and can be calculated as follow:

\begin{equation}\label{r_ISCO_01}
    \left\{ {\begin{array}{*{20}{c}}
    {{\partial _r}{U_{eff}} = {\partial _\theta }{U_{eff}} = 0}\\
    {}\\
    {\partial _r^2{U_{eff}} = \partial _\theta ^2{U_{eff}} = 0}
    \end{array}} \right. \Rightarrow r = {r_{ISCO}}\,.
\end{equation}

So,

\begin{equation}\label{r_ISCO_02}
{\left. {\partial _r^2{U_{eff}}} \right|_{r = {r_{ISCO}}}} = {\left. {\frac{{{{\tilde E}^2}\partial _r^2{g_{\phi \phi }} + 2\tilde E\tilde L\partial _r^2{g_{t\phi }} + {{\tilde L}^2}\partial _r^2{g_{tt}} - \partial _r^2\left( {g_{t\phi }^2 - {g_{tt}}{g_{\phi \phi }}} \right)}}{{g_{t\phi }^2 - {g_{tt}}{g_{\phi \phi }}}}} \right|_{r = {r_{ISCO}}}} = 0{\mkern 1mu} \,.
\end{equation}

In fact, the smallest radius of the circular orbit during which the test particle rotates statically around a black hole is called the innermost stable circular orbit (ISCO). So if we want to formulate the ISCO in the radial or vertical component, we have to consider the following condition for the effective potential:

\begin{equation}\label{ISCO_component_01}
    r = {r_{ISCO}} \Leftrightarrow \partial _r^2{U_{eff}} = 0 \Rightarrow \left\{ {\begin{array}{*{20}{c}}
    {\partial _r^2{U_{eff}} > 0\; \textrm{or}\;\partial _\theta ^2{U_{eff}} > 0 \to \textrm{The orbit is radially (vertically) stable}}\\
    {}\\
    {\partial _r^2{U_{eff}} < 0\; \textrm{or}\;\partial _\theta ^2{U_{eff}} < 0 \to \textrm{The orbit is radially (vertically) unstable}}
    \end{array}} \right.
\end{equation}

According to the definition of the ISCO, we find that it’s a relativistic phenomenon. From the point of view of Newtonian mechanics, circular orbits are formed in all radii, but in GR, circular orbits are formed in specific radii. It should be noted that according to the effective potential formula in the general relativity, the effective potential has extreme points in radial and vertical components and the local minimum of the effective potential is location of the ISCO. Figure~\eqref{U_eff_03} shows the effective potential as a function of the radial coordinate for the Kerr black hole and for the effective rotating charged spacetime considered in this work, with the spin parameter fixed to $a = 0.9$. The dependence of the effective potential on different values of the effective curvature parameter $\alpha$ and the charge parameter $Q$ is also illustrated.

The figure shows that increasing either $\alpha$ or $Q$ shifts the minimum of the effective potential toward smaller radii, corresponding to a reduction of the innermost stable circular orbit (ISCO). Modifications of the spacetime geometry associated with the effective curvature parameter alter the slope of the effective potential near the ISCO, allowing stable circular orbits closer to the black hole. In contrast, the presence of charge modifies the slope of the effective potential through electrostatic repulsion, thereby influencing the stability and location of circular orbits.

\subsection{Accretion disk energy efficiency}

We can define the maximum energy extracted from the accretion disk as the efficiency of the particle falling into the black hole. One of the energy transfer mechanisms for the particle that accrete into the black hole is radiation. So, the energy-momentum of the accreted particles is transferred outward by converting into electromagnetic radiation. Since the ISCO is the inner most stable circular orbit, we can obtain the rate of radiation of the particle energy-momentum from the surface of the accretion disk. This parameter is known as the energy efficiency and can be computed with the following equation:

\begin{equation}\label{efficiency_01}
  \eta  = 1 - {E_{ISCO}}\,.
\end{equation}

Observations of accretion processes around compact objects with and without a solid surface---such as neutron stars and black holes---indicate the existence of different accretion efficiencies~\cite{56}. For this reason, the determination of the innermost stable circular orbit (ISCO) parameters, namely the specific energy $E_{\mathrm{ISCO}}$, specific angular momentum $L_{\mathrm{ISCO}}$, and angular velocity $\Omega_{\mathrm{ISCO}}$, plays a crucial role in the analysis of accretion disk efficiency. Using the relations derived in the previous sections, we compute these quantities and summarize the results in Table~\ref{Table1}.

Table~\ref{Table1} presents the ISCO parameters for test particles in the Kerr spacetime and in the effective rotating charged spacetime considered in this work, with the spin parameter fixed to $a = 0.9$. The results show that increasing the charge parameter $Q$ and the effective curvature parameter $\alpha$ leads to a decrease in both the ISCO radius and the accretion efficiency. As the values of $Q$ and $\alpha$ increase, stable circular orbits are allowed closer to the black hole, while the corresponding accretion efficiency is reduced.

This behavior reflects modifications of the spacetime geometry in the strong-field region, which alter particle dynamics near the ISCO. Although particles can remain stable at smaller radii, they release less energy before plunging into the black hole compared to the Kerr case. These results highlight the sensitivity of accretion efficiency to effective geometric deformations and charge effects in high-spin configurations.

\begin{table}
\centering
\begin{tblr}{
  cell{1}{1} = {c=4}{},
}
Circular Motion Parameter for a=0.9 &       &         &        \\
\hline
$Q$                                   & ${\alpha}$ & $r_{ISCO}$    & $\eta $   \\
\hline
0                                   & 0     & 4.5     & 0.1942 \\
0.5                                 & 0.1   & 2.33165 & 0.0655 \\
0.5                                 & 0.01  & 2.43612 & 0.0750 \\
0.05                                & 0.1   & 2.33612 & 0.0646 \\
0.05                                & 0.01  & 2.5213  & 0.0657 
\end{tblr}
\caption{Numerical values of the ISCO radius $r_{\mathrm{ISCO}}$ and accretion efficiency $\eta$ for test particles in the effective rotating charged spacetime for different values of the effective curvature parameter $\alpha$ and the charge parameter $Q$, with the spin fixed to $a = 0.9$. Results are compared with the Kerr case.
}\label{Table1}
\end{table}

\subsection{Accretion disk radiation flux}

In the previous section, we discussed the energy-momentum transfer of particles in the accretion disk. One of the methods of energy-momentum transfer in the accretion process is that particles lose their energy-momentum and move towards the inner orbits of the accretion disk and accrete into the black hole. In this process, the energy-momentum transfer of particles emits radiation towards the outer layers of the orbits \cite{57}. Therefore, using the law of conservation of energy-momentum, we write the following equation for the radiative energy flux of the accretion disk \cite{58}:

\begin{equation}\label{radiation_flux_01}
  K(r) =  - \frac{{\dot M{\Omega _{,r}}}}{{4\pi \sqrt { - g} {{(\tilde E - \tilde L\Omega )}^2}}}\int_{{r_{ISCO}}}^r {(\tilde E - \tilde L\Omega )} {{\tilde L}_{,r}}dr \,.
\end{equation}

The accretion disk radiation flux $K(r)$, defined as the power of the radiation per unit area, is directly related to the energy dissipation mechanisms in the disk \cite{32}. This radiation flux is generated by internal turbulent viscosity, which converts the mechanical energy of rotation into internal heat \cite{33} and this heat, radiates into space through radiative (and sometimes convective) processes \cite{34}. Kerr black holes have a ring singularity and a smaller event horizon than Schwarzschild black holes \cite{35}. Most importantly, the rotation of the black hole causes the space-time around it to be dragged (frame-dragging effect), shifting the position of $r_{ISCO}$ towards the event horizon \cite{36}.

\begin{figure}[htb]
\centering
\subfloat[\label{K1} Accretion disk radiation flux in the Kerr spacetime and the effective rotating and rotating charged geometries.]{\includegraphics[width=0.775\textwidth]{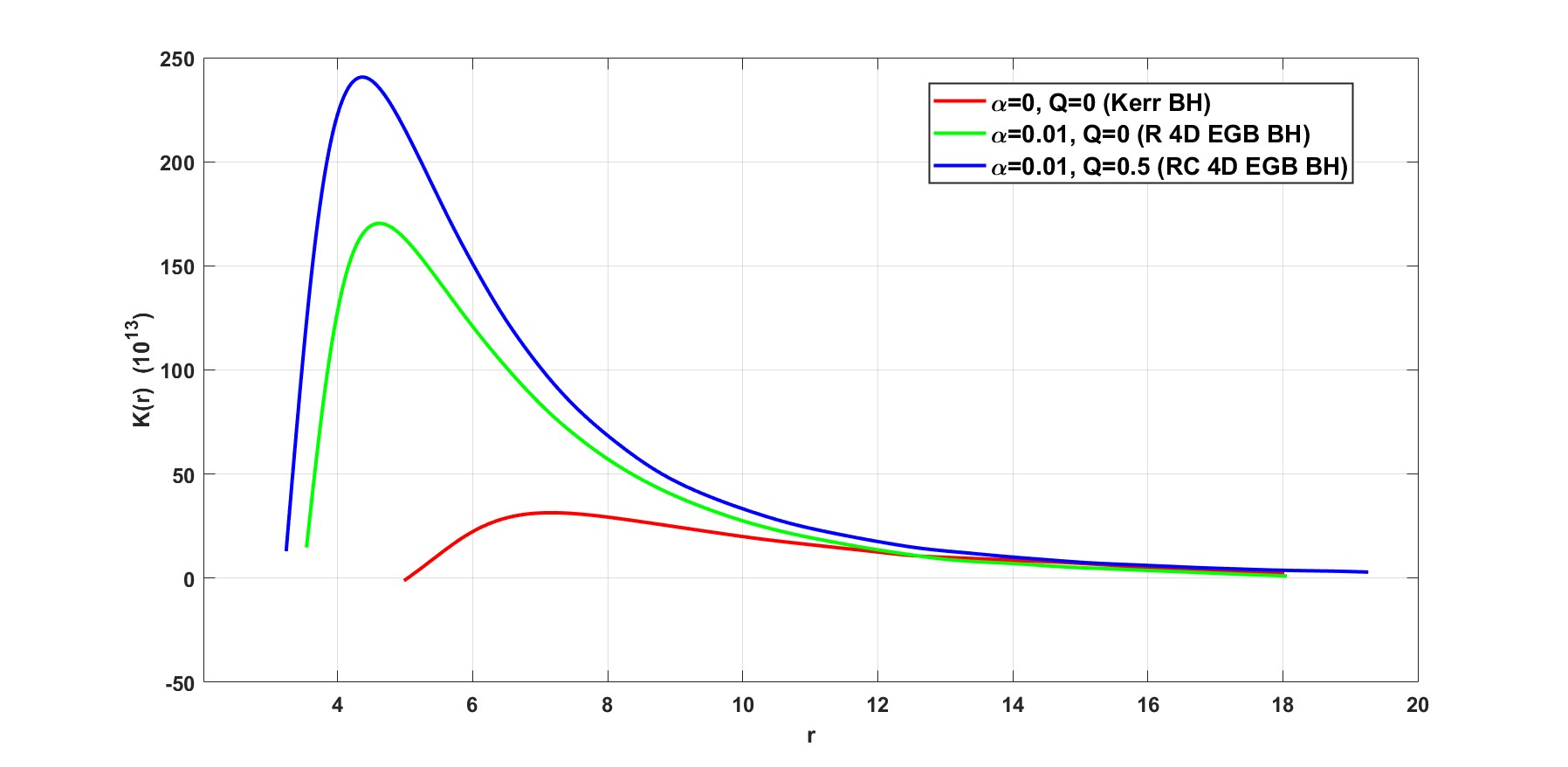}}
\,\,\,
\subfloat[\label{K2} Accretion disk radiation flux for different values of $\alpha$ and $Q$ in the effective rotating charged spacetime.
]{\includegraphics[width=0.775\textwidth]{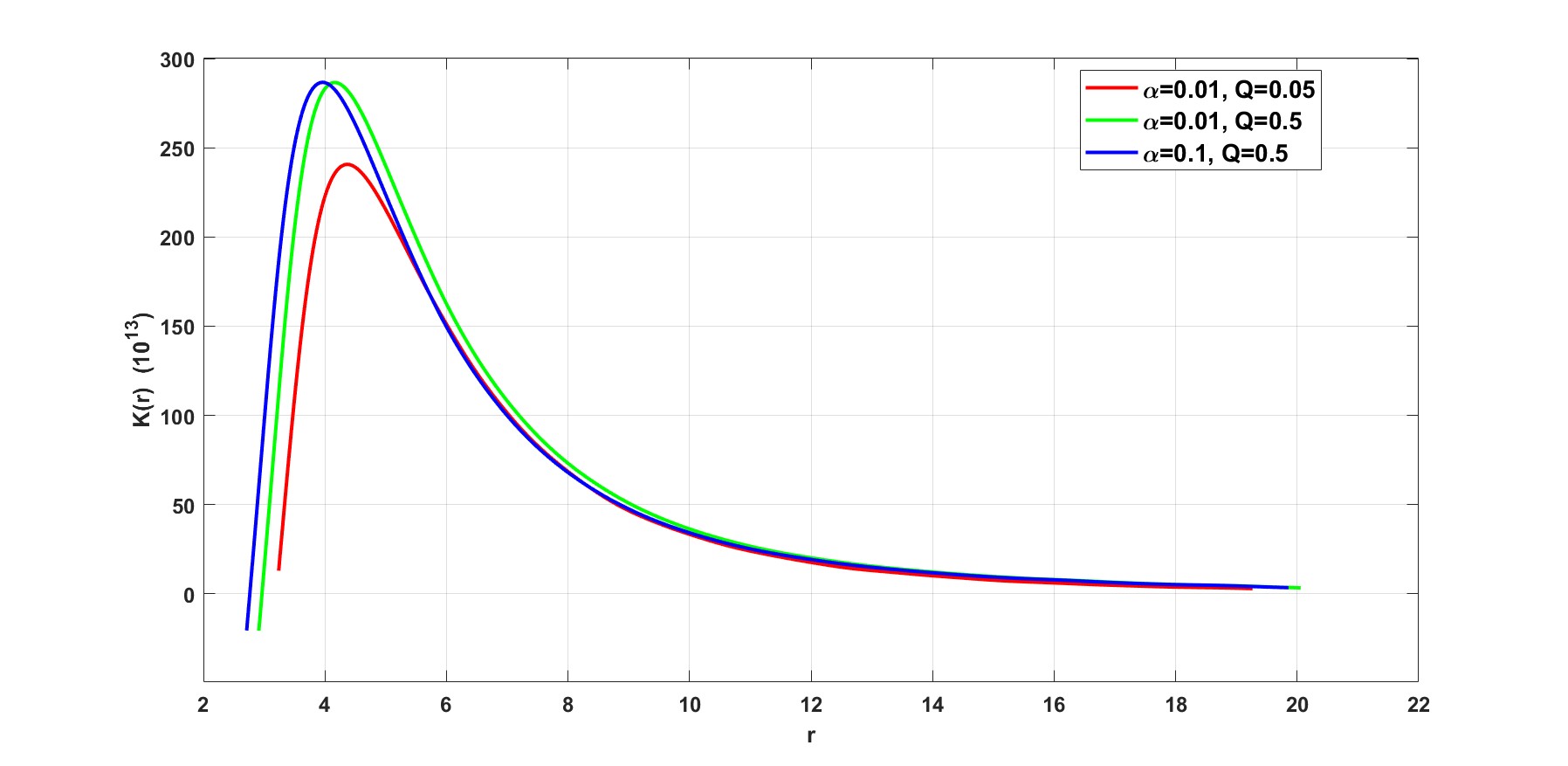}}
\caption{\label{K_02}\small{\emph{Accretion disk radiation flux for the Kerr black hole and the effective rotating charged spacetime, shown for different values of $\alpha$ and $Q$ with $a = 0.9$.
}}}
\end{figure}

Figure~\eqref{K_02} shows the radiation flux profile of the accretion disk around the effective rotating charged spacetime considered in this work. As illustrated in the figure, increasing the effective curvature parameter leads to a significant enhancement of the radiation flux, particularly in the inner region of the accretion disk. This behavior indicates that geometric modifications of the spacetime can concentrate energy release closer to the black hole, potentially giving rise to higher-energy emission such as X-rays.

In contrast, the presence of a positive charge parameter reduces the radiation flux. This reduction can be attributed to electrostatic repulsion, which lowers the effective matter density in the accretion disk and weakens the energy dissipation rate. The combined influence of effective curvature-induced geometric deformation and charge effects therefore plays a crucial role in shaping the radiative properties of the accretion disk.

This shifting increases the efficiency of mass-to-energy conversion and significantly affects the energy distribution and the radiation flux of the accretion disk. The radiation flux around Kerr black holes typically peaks at smaller radii than Schwarzschild black holes due to the small $r_{_{ISCO}}$ and frame stretching \cite{37}. Similarly, electric charge not only affects the geometry of space-time (changing event horizons and $r_{_{ISCO}}$), but also directly interacts with the electromagnetic fields and plasma in the accretion disk \cite{38}. These interactions can activate additional energy dissipation mechanisms (such as Joule dissipation) and potentially significantly change the radiation flux $K(r)$ \cite{39}.

\subsection{Temperature of the accretion disk}

The accretion disk temperature is one of the most important observable and calculable quantities in the study of accretion processes around compact objects, particularly black holes. The temperature of the disk is directly related to the radiation flux $K(r)$, and its radial distribution plays a central role in shaping the observed electromagnetic spectrum. A detailed understanding of the disk temperature profile is therefore essential not only for interpreting astrophysical observations, but also for probing how deviations from the Kerr geometry may influence accretion phenomena~\cite{40}.

In the simplest approximation, the accretion disk can be modeled as an ideal blackbody emitter. Within this framework, the radiation flux is directly related to the surface temperature of the disk at each radius~\cite{41}. This relationship is described by the Stefan--Boltzmann law for blackbody radiation, which can be written as follows:

\begin{equation}\label{Temperature_01}
  K(r) = {\sigma _{_{SB}}}T^{4} (r)\,.
\end{equation}

\begin{figure}[htb]
\centering
\subfloat[\label{T1} Accretion disk temperature in the Kerr spacetime and the effective rotating and rotating charged geometries.]{\includegraphics[width=0.775\textwidth]{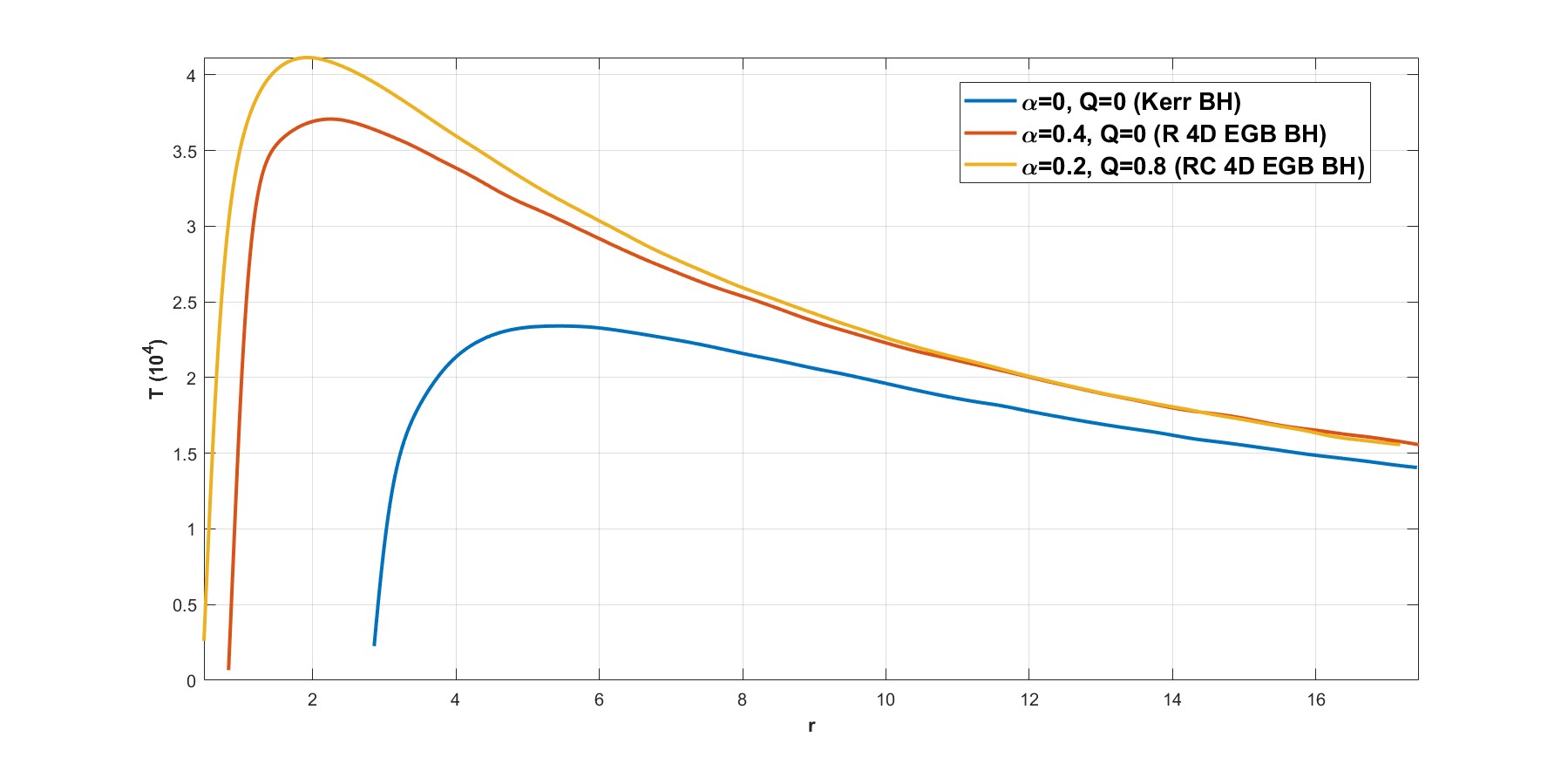}}
\,\,\,
\subfloat[\label{T2} Accretion disk temperature for different values of $\alpha$ and $Q$ in the effective rotating charged spacetime.
]{\includegraphics[width=0.775\textwidth]{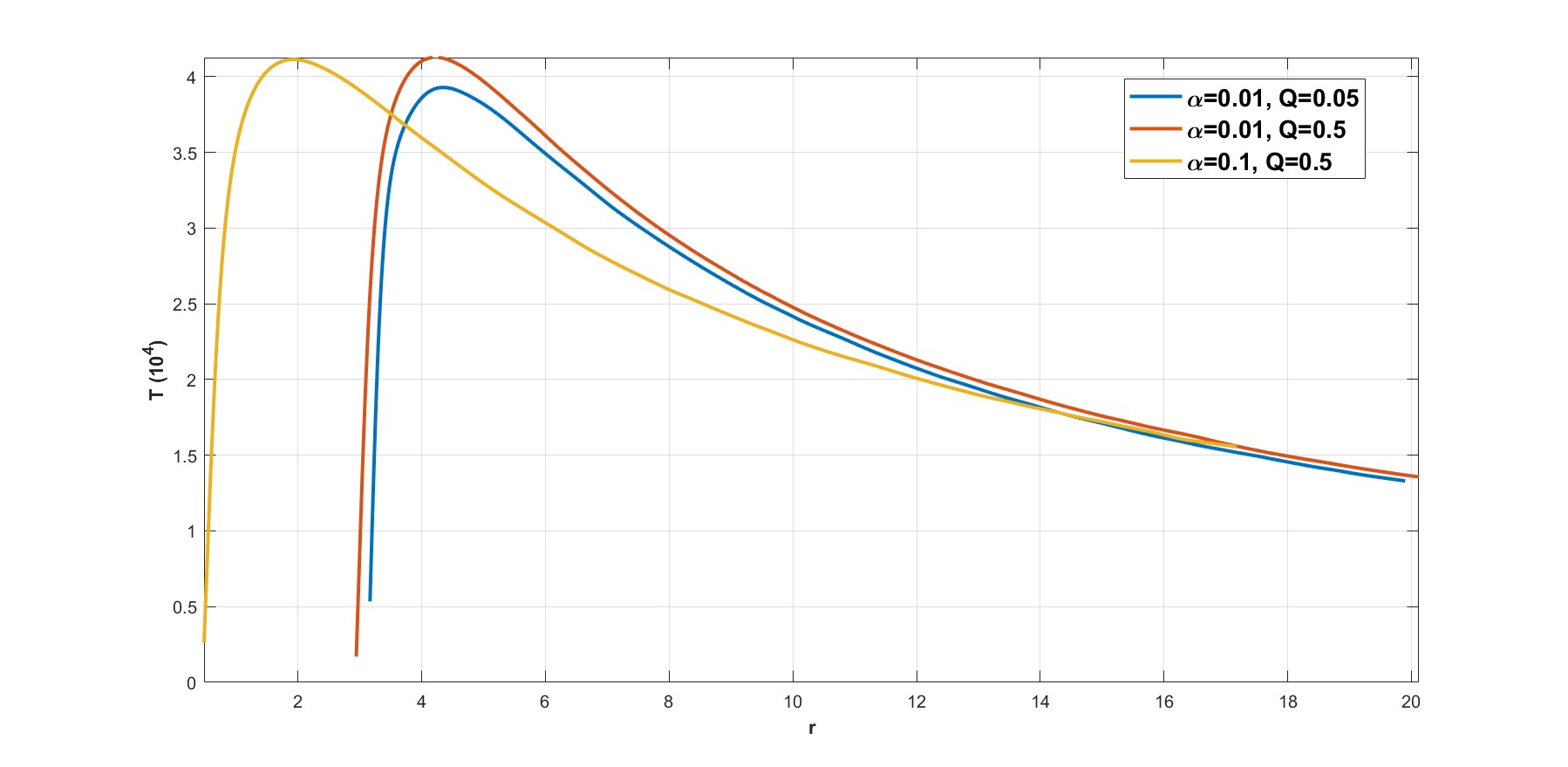}}
\caption{\label{Temperature_02}\small{\emph{Accretion disk temperature for the Kerr black hole and the effective rotating charged spacetime, shown for different values of $\alpha$ and $Q$ with $a = 0.9$.
}}}
\end{figure}

Figure~\eqref{Temperature_02} shows the temperature profile of the accretion disk around the effective rotating charged spacetime considered in this work. As expected, the disk temperature decreases with increasing radial distance from the black hole, since matter emits less energy at larger radii. Consistent with the behavior of other accretion disk observables, increasing the effective curvature parameter leads to an enhancement of the disk temperature, particularly in the inner region, while increasing the charge parameter results in a reduction of the temperature due to electrostatic repulsion and reduced energy dissipation.

The figure demonstrates that the temperature distribution within the accretion disk is strongly influenced by the black hole parameters---such as spin and charge---as well as by effective geometric deformations of the spacetime. Compared to the Kerr case, the effective rotating charged spacetime exhibits quantitative deviations in the temperature profile, reflecting modifications of the spacetime geometry in the strong-field regime.

Since the temperature in the accretion disk is not  homogeneous and increses while one decreases the radius, so the disk produces a multicolor blackbody spectrum. The final spectrum of the disk is the sum (integral) of blackbody spectra with different temperatures across the radii of the disk \cite{42,43}. The temperature distribution provides crucial information about how energy is distributed and dissipated within the disk. The temperature is affected by the black hole parameters. For example, as $r_{_{ISCO}}$ becomes smaller, matter accretes into regions with stronger gravitational fields and higher temperatures \cite{60,61}. This leads to an increase in temperature and a shift of the emission peak to higher energies (e.g., X-rays) \cite{44,45}. Optical spectra, and X-ray spectra of accretion disks around binary black holes and AGNs are probed \cite{62}. The shape of these spectra (especially the X-ray component often attributed to the accretion disk) can be compared with computationally based spectral models $T(r)$. This comparison allows the derivation of important physical parameters such as mass, spin, accretion rate, etc of the black holes \cite{23}.

\subsection{Luminosity of the accretion disk}

Luminosity of an accretion disk is one of the most fundamental and measurable quantities in astrophysics. This quantity represents the total radiation power emitted from the disk surface in all directions and across the electromagnetic spectrum. Luminosity is directly related to the radiation flux $K(r)$ and temperature $T(r)$, which were discussed in the previous subsections. Studying the luminosity and its dependence on the black hole parameters provides a powerful tool for understanding the efficiency of the accretion process \cite{40,41}. Differential luminosity is the total luminosity that depends on a specific variable. In fact, differential luminosity tells us how luminosity would received to an infinity distant observer. If we assume the accretion disk is optically thick, energy dissipation may manifest itself as luminosity at surface of the accretion disk. Hence we can obtain the relationship between differential luminosity and radiation flux as follows:

\begin{equation}\label{Luminosity_01}
  \frac{{d{{\cal L}_\infty }}}{{d\ln r}} = 4\pi r {\tilde E} \sqrt g K\left( r \right) \,.
\end{equation}

\begin{figure}[htb]
\centering
\subfloat[\label{lnu1} Differential luminosity of the accretion disk in the Kerr spacetime and the effective rotating and rotating charged geometries.]{\includegraphics[width=0.775\textwidth]{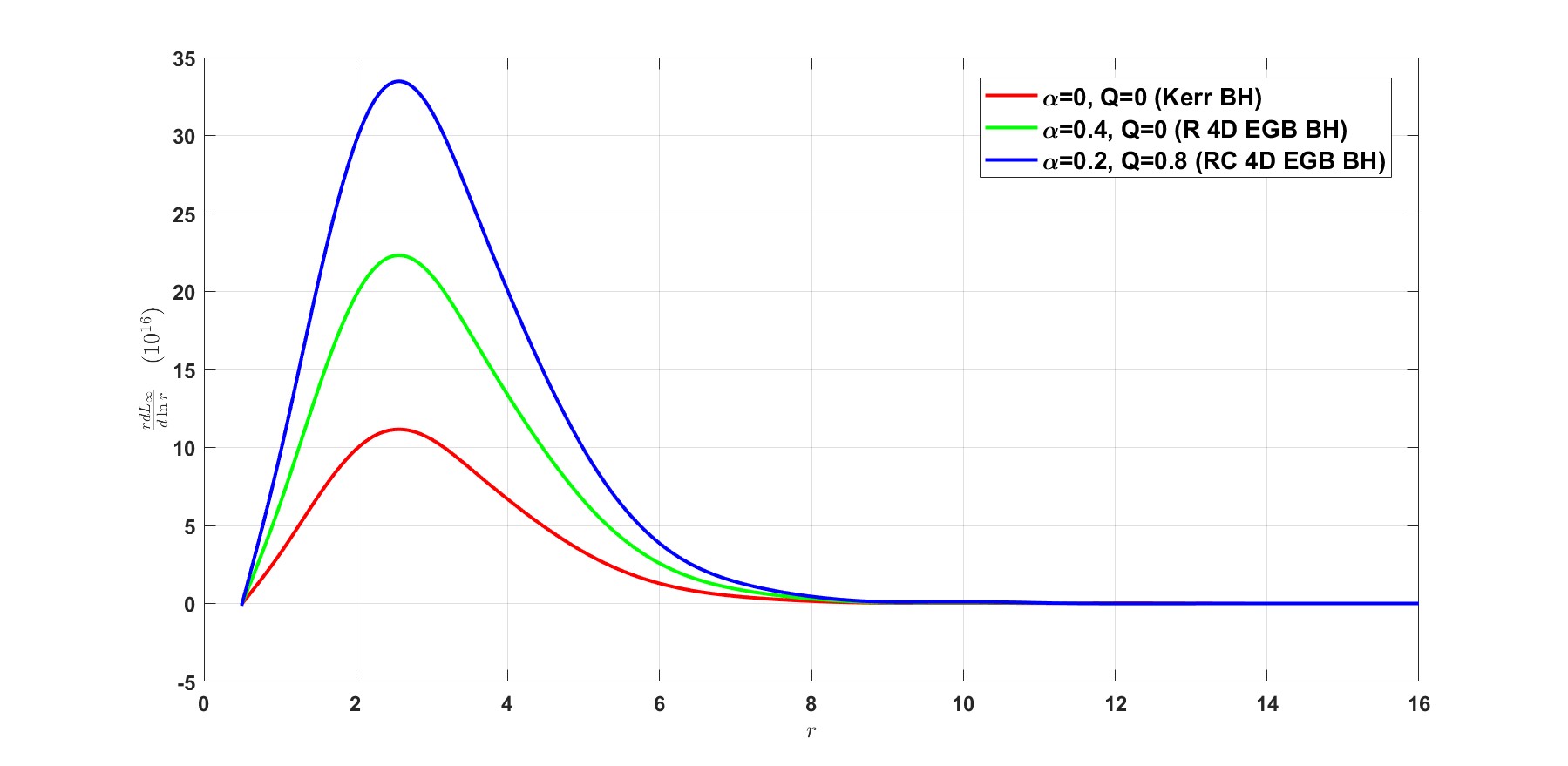}}
\,\,\,
\subfloat[\label{lnu2} Differential luminosity of the accretion disk for different values of $\alpha$ and $Q$ in the effective rotating charged spacetime.]{\includegraphics[width=0.775\textwidth]{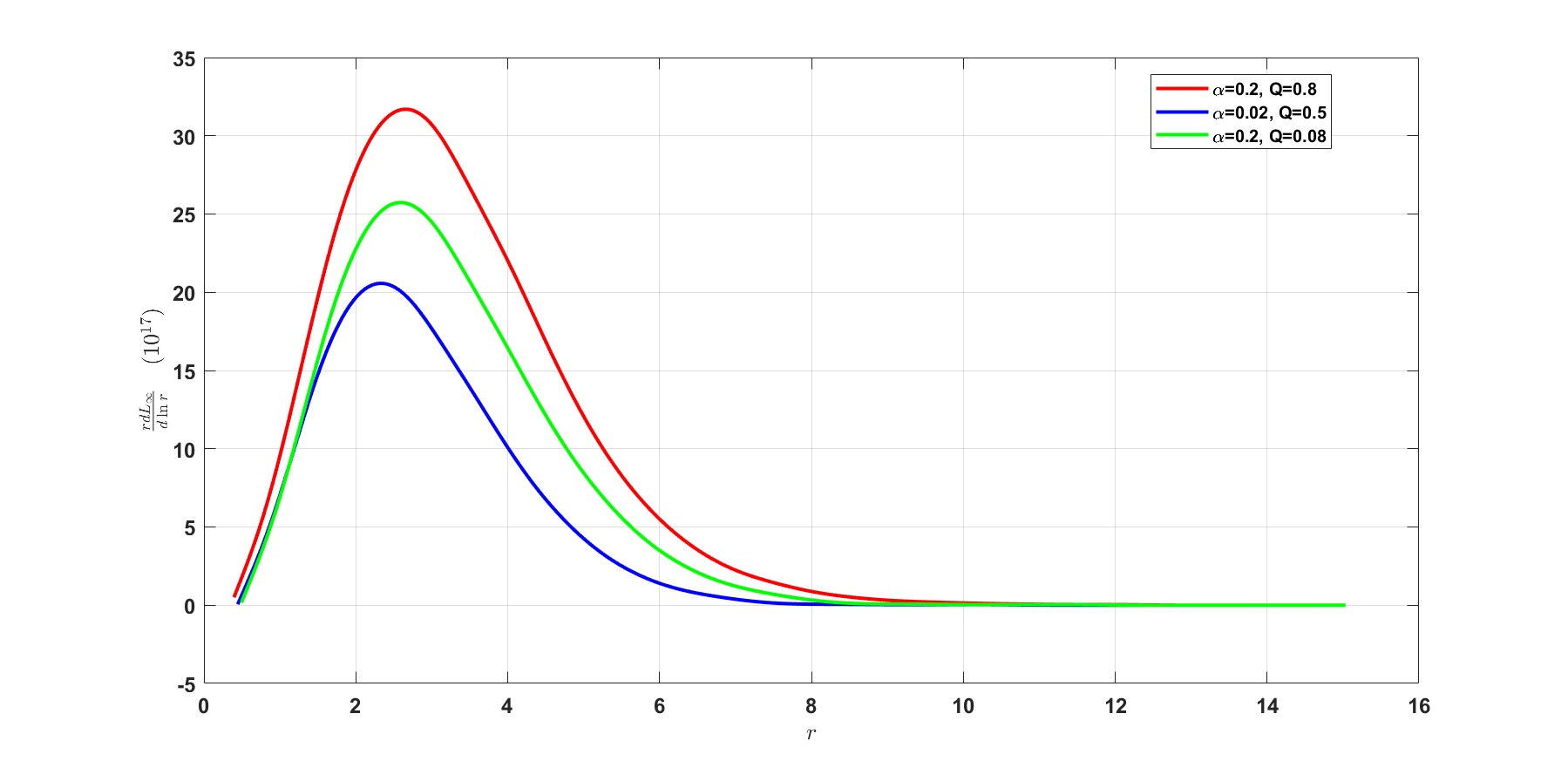}}
\caption{\label{Luminosity_02}\small{\emph{Figure~(\ref{lnu1}) shows the differential luminosity of the accretion disk for the Kerr black hole and the effective rotating charged spacetime, while Fig.~(\ref{lnu2}) illustrates its dependence on different values of $\alpha$ and $Q$ for $a = 0.9$. Increasing $\alpha$ enhances the differential luminosity, whereas increasing $Q$ leads to a reduction.}}}
\end{figure}

However, to compare the strengths of different accretion sources, we use the accretion disk differential luminosity for various scenario, which helps to understand the inherent similarities and differences of the astrophysical systems \cite{41}. Unlike the flux or temperature, which are local distributions and their extraction requires angular resolution or high-precision spectroscopy, the total luminosity can be estimated with good accuracy even for far distant sources \cite{46}. Figure~\eqref{Luminosity_02} shows the differential luminosity of the accretion disk around the effective rotating charged spacetime considered in this work. The shift of the innermost stable circular orbit (ISCO) toward smaller radii allows matter to accumulate closer to the black hole, where gravitational effects are stronger and energy release is more efficient. As a result, increasing the effective curvature parameter leads to an enhancement of the differential luminosity of the accretion disk.

In contrast, increasing the charge parameter reduces the disk luminosity, since electrostatic repulsion lowers the effective matter density and weakens the rate of energy dissipation. Although energy transfer within the disk continues as long as particles accrete onto the black hole, the differential luminosity cannot grow without bound. In spherical accretion flows, the energy–momentum carried by radiation experiences a force proportional to the momentum flux density, which limits the radiative output \cite{63}.

\section{Rotational Effects, Centrifugal Force Reversal, and Equipotential Structures in Accretion Disk Dynamics}\label{RFoP}

Beyond circular geodesics and radiative properties, the internal structure and stability of accretion disks are strongly influenced by rotational effects in the underlying spacetime. In particular, centrifugal force reversal and the topology of equipotential (von Zeipel) surfaces play a key role in determining the behavior of fluid elements, angular momentum transport, and the inner edge of the disk. In this section, we analyze these effects in the effective rotating spacetime considered here and discuss their implications for accretion disk structure.

The effects of increasing the rotation radius for the local outward direction are related to the dynamical features of rotation. However, in strong gravitational fields, these effects are anomalous. Radius of gyration in Newtonian theory, the radius of the circular path of a point-like particle with mass and angular momentum, is determined to be related to the angular momentum of a rigid body \cite{107}. Therefore, the radius of gyration comes from the following equation:

\begin{equation}\label{r_Von1}
  \tilde r = {\left( {\frac{J}{{M\Omega }}} \right)^{\frac{1}{2}}} \,,
\end{equation}
where $J$ is the angular momentum, $M$ is mass, and $\Omega$ is the angular velocity.

Obviously, for a point particle moving in a circular path, the value of $\tilde r$ will be the radius of a circle. In GR, the following two items are the two standard measures for measuring the radius of the mentioned circular orbit.

\begin{itemize}
  \item Circumferential radius (specific radius of the circumference of the circle divided by $2\pi$)
  \item	The radial proper distance
\end{itemize}

If we generalize these two criteria for a point particle moving in a circular path in GR, the above equation is obtained as follows:

\begin{equation}\label{r_Von2}
  \tilde r = {\left( {\frac{{\cal L}}{{{\cal E}{\rm{\Omega }}}}} \right)^{\frac{1}{2}}} \,,
\end{equation}
where ${\cal L}$ is the angular momentum of the particle and ${\cal E}$ is its energy.

Abramovich et al. suggested that $\tilde r$ is an appropriate quantity to discuss the effects of increasing the radius of gyration for the local outward direction \cite{23}. The top surfaces $\tilde r$ were introduced as von Zeipel cylinders, which is a new concept for analyzing the behavior of rotating fluid in different space-times. These surfaces are usually referred to as surfaces with constant $\frac{\ell }{{\rm{\Omega }}}$ rate ($\ell  = \frac{{\cal L}}{{\cal E}}$ is the characteristic angular momentum). But with such a new definition, we can analyze them as surfaces with constant radius of gyration \cite{30}.
In Newtonian mechanics, these surfaces are usually cylindrical. But their topology in GR is determined by the curvature of space-time and they may not be cylindrical \cite{64}. Awareness of this issue and its consequences is an important key to understanding rotational effects in GR. Here we need to specify the direction of the inward/outward flows. As we said before, we have two standard criteria \cite{65}.

\begin{itemize}
  \item Peripheral radius perspective: The direction of decreasing/increasing the inward/outward flow is determined by increasing the circumference of the concentric circles.
  \item Radial special distance perspective: The direction of decreasing/increasing the inward/outward flow is determined by increasing the value of r (the concept of von Zippel cylinders).
\end{itemize}

After specifying the direction of increase/decrease of the inward/outward flows, we need to generalize the definition of centrifugal force to GR. For this purpose, we examine the circular motion for a static and axially symmetric space-time (it has two killing vectors: timelike ${\eta ^\alpha }$ and spacelike ${\xi ^\alpha }$). In this case, the appropriate peripheral radius is formulated as $r = \sqrt {{\xi _\alpha }{\xi ^\alpha }}$ and the vector ${\nabla _\alpha }r$, where $r$ is defined as the outgoing direction. The general geodesic of motion for this state is formulated as follows:

\begin{equation}\label{r_Von3}
  \tilde r = \sqrt {\frac{{\cal L}}{{{\cal E}{\rm{\Omega }}}}}  = \sqrt { - \frac{{\left( {{\xi _\alpha }{\xi ^\alpha }} \right)}}{{\left( {{\eta _\alpha }{\eta ^\alpha }} \right)}}}  = \frac{{ - {g_{\phi \phi }}}}{{\sqrt {g_{t\phi }^2 - {g_{tt}}{g_{\phi \phi }}} }} \,,
\end{equation}
where ${\cal E} =  - {\eta ^\alpha }{v_\alpha }$ (energy), ${\cal L} = {\xi ^\alpha }{v_\alpha }$ (angular momentum) and ${v^\alpha }$ is the four velocity vectors, and the vector ${\nabla _\alpha}r$ is defined as the local outward direction. It should be noted that $\tilde r$ is also called the Von Zeipel radius.

The four velocity vector of a test particle with rest mass in circular motion is defined as follows:

\begin{equation}\label{v_a1}
  {v^\alpha } = \frac{{{\eta ^\alpha } + \Omega {\xi ^\alpha }}}{{\sqrt { - \left[ {\left( {{\eta _\alpha }{\eta ^\alpha }} \right) + {\Omega ^2}\left( {{\xi _\alpha }{\xi ^\alpha }} \right)} \right]} }} \,.
\end{equation}

This definition of four velocity vector is valid in the inertial frame. If we consider a stationary observer, we can visualize the four velocity vector into two parts of the stationary and three-space frames. As a result, orbital speed in three-space is calculated as follows:

\begin{equation}\label{v_tlde1}
  \tilde v = \frac{{{\rm{\Omega }}\tilde r}}{{\sqrt {1 - {{\rm{\Omega }}^2}{{\tilde r}^2}} }} \,.
\end{equation}

Therefore, the acceleration (${a_\alpha } \equiv {v^\beta }{\nabla _\beta }{v^\alpha }$) is obtained as follows.

\begin{equation}\label{a_a1}
  {a_\alpha } = \frac{1}{2}{\nabla _\alpha }\ln \left[ { - {\eta _\alpha} {\eta ^\alpha} } \right] - \frac{{{m_0}{{\tilde v}^\alpha }}}{{\tilde r}}{\nabla _\alpha }\tilde r \,.
\end{equation}

We know that the circular motion of the test particle occurs when the gravitational force and the centrifugal forces are equal. The question that arises here is which term of the acceleration is related to the gravitational force and which term is related to the centrifugal force.
Abramovich et al. believes that the gravitational force is independent of speed. As a result, the first term is related to gravity \cite{23}. Finally, the centrifugal force of a single ball with rest mass ${m_0}$ is as follows:

\begin{equation}\label{F01}
  {F_0} = \frac{{{m_0}{{\tilde v}^\alpha }}}{{\tilde r}}{\nabla _\alpha }\tilde r \,.
\end{equation}

This equation states that the centrifugal force is always localized in the outgoing direction since the centrifugal force is proportional to ${\nabla _\alpha }\tilde r$. Bini et al. argued ``Is it worth introducing the centripetal force into general relativity?'' and ``if we want to give a definition of centrifugal force, its direction should be towards the outside of the axis of rotation'' and in this way they questioned the concept of centrifugal force \cite{68,66,67}. It should be mentioned that in this definition, reversing the direction of the light-like geodesics does not change the centripetal force and still describes the dynamic effects of the rotation.

The equipotential surfaces of the photon movement exactly correspond to the top surfaces of $\tilde r$ (Von Zeipel cylinders) and the direction of the centripetal force ${F_\alpha}$ is in the direction of decreasing ${V_{_{eff}}}$. Photon's circular orbits appear in the effective potential and have upper and lower bounds. The upper limit corresponds to unstable orbits and the lower limit corresponds to the innermost stable circular orbit (ISCO). Therefore, the centrifugal force is zero for circular photon modes and in the polar plane it is always between unstable orbits and stable circular orbit (ISCO).
According to the definition of $\tilde r$ from the Eq. \eqref{r_Von3} and the metric \eqref{metric_01}, the following equation is obtained for $\tilde r$, $\tilde v$ and ${F_0}$ respectively:

\begin{equation}\label{r_Von4}
  \tilde r = \frac{{ - {a^2} + \Delta }}{{{r^2}\sqrt {\frac{{{a^2}\left( {4{{\left( {{a^2} + {r^2} - \Delta } \right)}^2} - \left( {{a^2} - \Delta } \right)\left( {{{\left( {{a^2} + {r^2}} \right)}^2} - \Delta } \right)} \right)}}{{{r^4}}}} }}
\end{equation}

\begin{figure}
  \centering
  \includegraphics[width=0.675\textwidth]{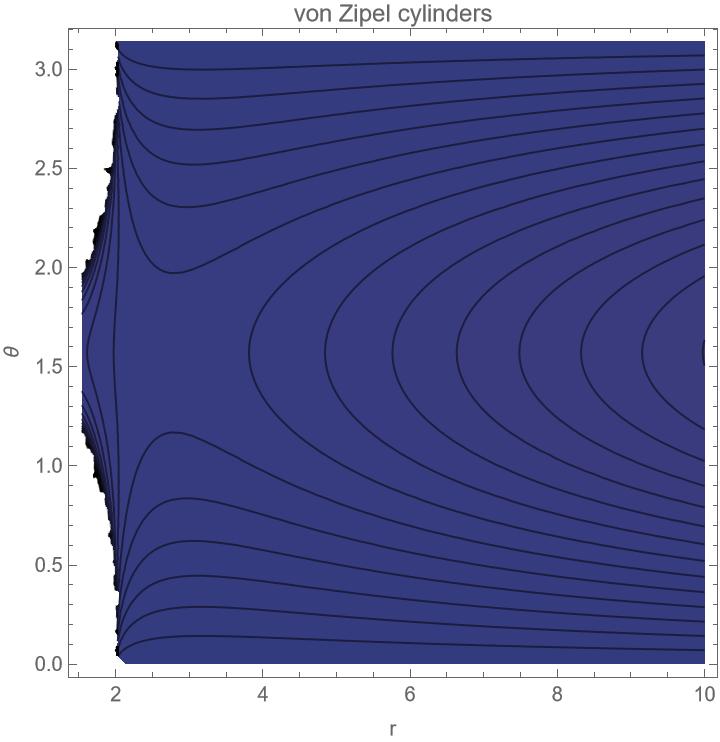}
  \caption{Von Zeipel cylinders for the effective rotating charged spacetime with $a = 0.9$, $\alpha = 0.01$, and $Q = 0.5$. The horizontal and vertical axes represent the radial coordinate and polar angle, respectively. Surfaces of constant von Zeipel radius $R$ are shown, illustrating their compression near the rotation axis and their behavior in the equatorial plane.
}\label{Vonzipple01}
\end{figure}

\begin{eqnarray*}
  \tilde v &=& \left( {{a^2} - \Delta } \right) \times \\
  && \left( {4{a^3} - 4a\Delta  + 2ar{\partial _r}\Delta  + {r^3}\sqrt {\frac{{{a^2}\left( {2{a^2} - 2\Delta  + r{\partial _r}\Delta } \right)\left( {2\left( {{a^4} + {r^4} + 2{a^2}\left( {2 + {r^2}} \right) + r\Delta  - 4\Delta } \right) - \left( { - 4 + r} \right)r{\partial _r}\Delta } \right)}}{{{r^6}}}} } \right) \\
  && \times \left[ {2\left( {{{\left( {{a^2} + {r^2}} \right)}^2} + r\Delta } \right) - {r^2}{\partial _r}\Delta } \right]{a^2}{r^2}\sqrt {\frac{{{a^2}\left( {4{{\left( {{a^2} + {r^2} - \Delta } \right)}^2} - \left( {{a^2} - \Delta } \right)\left( {{{\left( {{a^2} + {r^2}} \right)}^2} - \Delta } \right)} \right)}}{{{r^4}}}}  \\
  && \times \sqrt {1 + \frac{{{{\left( {{a^2} - \Delta } \right)}^2}{{\left( {4{a^3} - 4a\Delta  + 2ar{\partial _r}\Delta  + {r^3}B} \right)}^2}}}{{{a^6}\left( {\left( { - 4 + {a^2}} \right){{\left( {{a^2} + {r^2}} \right)}^2} - \left( {{a^2}\left( { - 7 + {a^2}} \right) + 2\left( { - 4 + {a^2}} \right){r^2} + {r^4}} \right)\Delta  - 3{\Delta ^2}} \right){{\left( { - 2\left( {{{\left( {{a^2} + {r^2}} \right)}^2} + r\Delta } \right) + {r^2}{\partial _r}\Delta } \right)}^2}}}}
\end{eqnarray*}

\begin{eqnarray*}
  {F_0} &=& - \frac{{\left[ {2M\left( {{a^2} - \Delta } \right)\left( {\left( { - 4 + {a^2}} \right)\left( {{a^2} + {r^2}} \right) - \left( { - 4 + {a^2} + {r^2}} \right)\Delta } \right)\left( {4{a^3} - 4a\Delta  + 2ar\Delta p + {r^3}B} \right)} \right]}}{{{a^2}r\sqrt {\frac{{{a^2}\left( {4{{\left( {{a^2} + {r^2} - \Delta } \right)}^2} - \left( {{a^2} - \Delta } \right)\left( {{{\left( {{a^2} + {r^2}} \right)}^2} - \Delta } \right)} \right)}}{{{r^4}}}} }} \\
  && \times \frac{{\left[ {2M\left( {{a^2} - \Delta } \right)\left( {\left( { - 4 + {a^2}} \right)\left( {{a^2} + {r^2}} \right) - \left( { - 4 + {a^2} + {r^2}} \right)\Delta } \right)\left( {4{a^3} - 4a\Delta  + 2ar\Delta p + {r^3}B} \right)} \right]}}{{\left[ {\left( { - 4 + {a^2}} \right){{\left( {{a^2} + {r^2}} \right)}^2} - \left( {{a^2}\left( { - 7 + {a^2}} \right) + 2\left( { - 4 + {a^2}} \right){r^2} + {r^4}} \right)\Delta  - 3{\Delta ^2}} \right]\left[ {2\left( {{{\left( {{a^2} + {r^2}} \right)}^2} + r\Delta } \right) - {r^2}\Delta p} \right]}} \\
  && \times \frac{{\left[ {2M\left( {{a^2} - \Delta } \right)\left( {\left( { - 4 + {a^2}} \right)\left( {{a^2} + {r^2}} \right) - \left( { - 4 + {a^2} + {r^2}} \right)\Delta } \right)\left( {4{a^3} - 4a\Delta  + 2ar\Delta p + {r^3}B} \right)} \right]}}{{\sqrt {1 + \frac{{{{\left( {{a^2} - \Delta } \right)}^2}{{\left( {4{a^3} - 4a\Delta  + 2ar{\partial _r}\Delta  + {r^3}B} \right)}^2}}}{{{a^6}\left( {\left( { - 4 + {a^2}} \right){{\left( {{a^2} + {r^2}} \right)}^2} - \left( {{a^2}\left( { - 7 + {a^2}} \right) + 2\left( { - 4 + {a^2}} \right){r^2} + {r^4}} \right)\Delta  - 3{\Delta ^2}} \right){{\left( { - 2\left( {{{\left( {{a^2} + {r^2}} \right)}^2} + r\Delta } \right) + {r^2}{\partial _r}\Delta } \right)}^2}}}} }}
\end{eqnarray*}
where by definition $B = \sqrt {\frac{1}{{{r^6}}}{a^2}\left( {2{a^2} - 2\Delta  + r{\partial _r}\Delta } \right)\left( {2\left( {{a^4} + {r^4} + 2{a^2}\left( {2 + {r^2}} \right) + r\Delta  - 4\Delta } \right) - \left( { - 4 + r} \right)r{\partial _r}\Delta } \right)}$ and $\Delta  = {r^2} + {a^2} + \frac{{{r^2}}}{{2\alpha }}\left[ {1 - \sqrt {1 + 4\alpha \left( {\frac{{2M}}{{{r^3}}} - \frac{{{Q^2}}}{{{r^4}}}} \right)} } \right]$.

Figure~\ref{Vonzipple01} shows the von Zeipel cylinders for the effective rotating charged spacetime considered in this work, with the parameters fixed to $a = 0.9$, $\alpha = 0.01$, and $Q = 0.5$. In this figure, the horizontal axis represents the radial coordinate, while the vertical axis corresponds to the polar angle, and surfaces of constant von Zeipel radius $R$ are plotted. As indicated by the figure, values of $\theta \simeq 1.5$ lie close to the rotation axis, where the von Zeipel radius asymptotically approaches $R \to 0$. Near the rotation axis, spacetime exhibits strong geometric compression, causing the von Zeipel cylinders to contract and particle trajectories to align toward the axis and vertical directions.

In this region, the von Zeipel cylinders are strongly influenced by the gravitational field, and the direction of the centrifugal force is perpendicular to the von Zeipel surfaces. The equatorial plane corresponds to $\theta = 1$, where the curvature effects are particularly pronounced. The figure also shows that strong-field spacetime curvature leads to inward bending of the von Zeipel surfaces, especially near the equatorial plane, accompanied by compression along the rotation axis.

\begin{figure}
  \centering
  \includegraphics[width=0.675\textwidth]{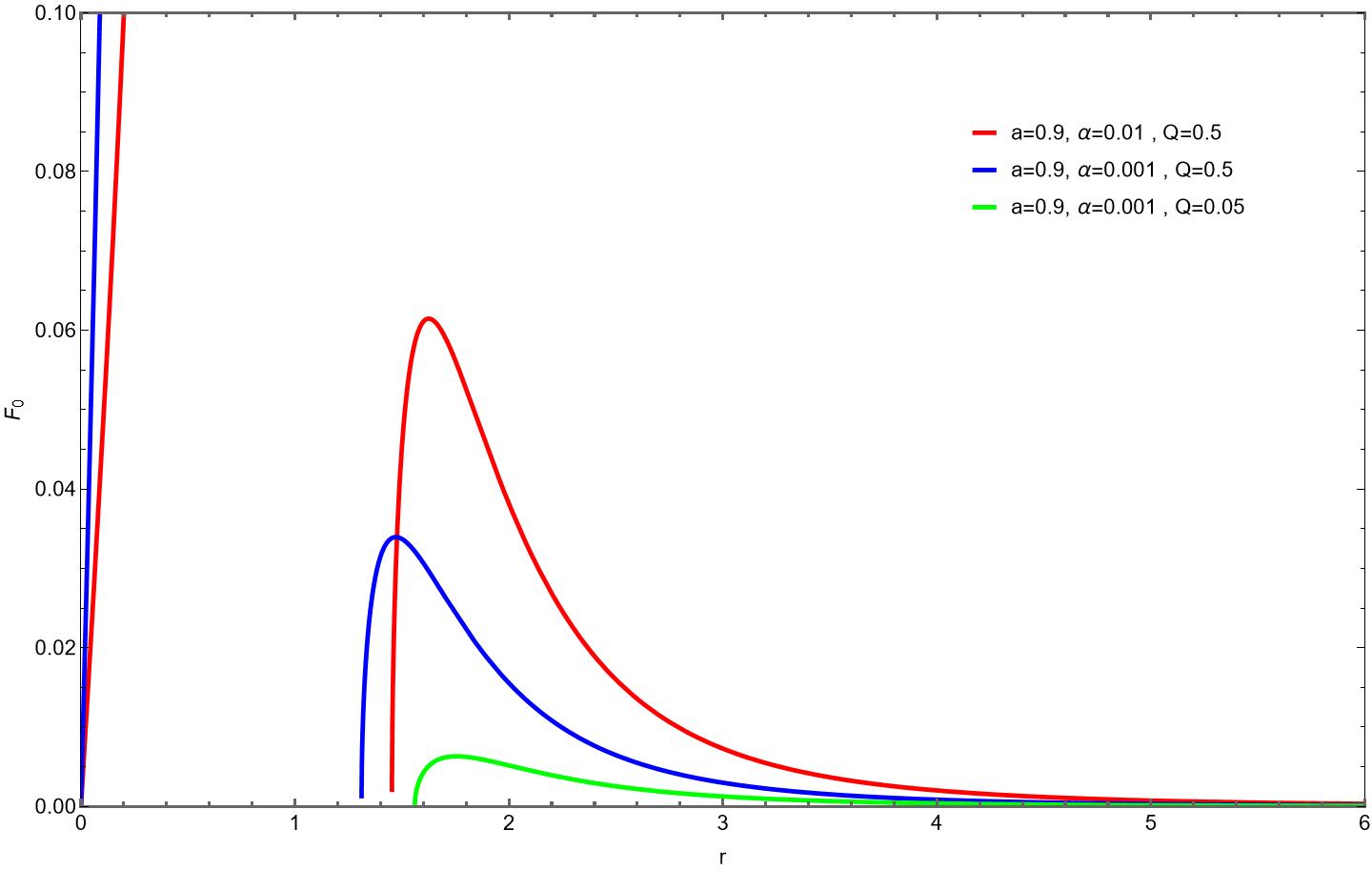}
  \caption{Centrifugal force for different values of $\alpha$ and $Q$ in the effective rotating charged spacetime with $a = 0.9$. The centrifugal force initially increases as particles approach the event horizon, reaches a maximum, and then decreases, exhibiting centrifugal force reversal in the strong-field region.}\label{Fc01}
\end{figure}

Figure~\ref{Fc01} illustrates the behavior of the centrifugal force experienced by particles as they approach the black hole. As expected from classical considerations, the centrifugal force initially increases as the radial distance decreases. However, after reaching a maximum value, it rapidly decreases and may even reverse direction in the vicinity of the event horizon. This phenomenon, known as centrifugal force reversal, arises due to the strong curvature of spacetime in the near-horizon region. The rotation of the black hole causes this reversal to occur at smaller radial distances. Increasing the effective curvature parameter $\alpha$ and the charge parameter $Q$ enhances the amplitude of the centrifugal force, although the radial location at which the reversal occurs varies depending on the parameter values.

The deformation of von Zeipel surfaces and the occurrence of centrifugal force reversal in the vicinity of the event horizon indicate significant modifications of the inner accretion flow. Such effects can influence disk thickness, stability against perturbations, and the transition from disk accretion to plunging motion. These features provide additional insight into how effective higher-curvature–inspired deformations of the spacetime geometry may alter the behavior of matter in the strong-field region of accretion disks.

The deviations from the Kerr geometry discussed in this work can, in principle, manifest in observable properties of accretion disks. In particular, modifications of the ISCO radius directly affect the inner edge of the disk and can lead to measurable changes in the high-energy portion of the emitted spectrum. Variations in the radiation flux and temperature profile influence the overall spectral shape and luminosity, especially in the X-ray band. These effects may be probed through observational techniques such as continuum-fitting and X-ray reflection spectroscopy using instruments like NuSTAR and XMM-Newton. In addition, alterations of the near-horizon spacetime structure may indirectly impact black hole imaging observations, such as those performed by the Event Horizon Telescope, by modifying the emission profile of the inner accretion flow. We emphasize, however, that the present analysis is primarily qualitative and aims to identify potential signatures of effective higher-curvature corrections. A detailed quantitative comparison with observational data is left for future work

\section{Summary and conclusions}\label{SaC}

In this work, we have investigated the structure and physical properties of thin accretion disks around a rotating charged black hole described by an effective higher-curvature--inspired spacetime, which represents a phenomenological deformation of the Kerr--Newman geometry. Adopting a purely kinematical approach, we analyzed test-particle motion and key accretion disk observables, including the specific energy, specific angular momentum, angular velocity, innermost stable circular orbit (ISCO), radiative efficiency, radiation flux, temperature profile, and differential luminosity. In addition, rotational effects in strong gravitational fields and the structure of von Zeipel (equipotential) surfaces were examined.

Our analysis shows that the effective curvature parameter $\alpha$ and the charge parameter $Q$ play a significant role in shaping particle dynamics and accretion disk properties. The numerical results and figures presented throughout the paper clearly demonstrate that increasing either $\alpha$ or $Q$ leads to a reduction in the specific energy and angular velocity of particles on circular orbits. This behavior reflects the modification of the spacetime geometry induced by higher-curvature--inspired corrections, together with the influence of electrostatic repulsion associated with the black hole charge.

We further find that the specific angular momentum decreases and that the ISCO radius shifts toward smaller values as $\alpha$ and $Q$ increase. These effects result in a reduction of the accretion efficiency, as confirmed by the numerical values reported in Table~\ref{Table1}, indicating that particles release less energy during their inward spiral toward the black hole. Such changes in the ISCO and efficiency directly affect the energy budget and emission properties of the accretion disk.

The radiation flux and temperature profiles of the accretion disk exhibit a complementary behavior: the effective curvature parameter $\alpha$ enhances both the radiation flux and the disk temperature, while the charge parameter $Q$ suppresses them. This suggests that stronger effective gravitational fields associated with higher-curvature deformations concentrate matter and energy release closer to the black hole, potentially enhancing high-energy emission such as X-rays. In contrast, the presence of charge reduces the effective matter density in the disk due to electrostatic repulsion, leading to diminished radiation output.

The differential luminosity of the accretion disk follows the same qualitative trends, with higher values associated with increasing $\alpha$ and lower values for larger $Q$. In parallel, the analysis of von Zeipel (equipotential) surfaces reveals noticeable deformations in the vicinity of the black hole, highlighting the strong influence of spacetime curvature and rotation on the structure of fluid equilibrium surfaces. These deformations become more pronounced as the effective curvature parameter increases.

Finally, our study of centrifugal force reversal near the event horizon shows that this purely relativistic phenomenon is sensitive to both $\alpha$ and $Q$. The presence of centrifugal force reversal provides further evidence of the extreme gravitational environment near the black hole and illustrates how effective geometric deformations can alter dynamical behavior in the strong-field regime.

Overall, our results demonstrate that effective higher-curvature--inspired deformations of rotating charged black hole spacetimes lead to observable deviations from the standard Kerr and Kerr--Newman predictions in accretion disk properties. While the present study does not rely on a specific fundamental gravitational theory, it highlights how phenomenological modifications of the spacetime geometry can leave imprints on accretion observables. Such effects may become accessible to future high-precision observations, including black hole imaging and spectral measurements from facilities such as the Event Horizon Telescope.

We conclude that thin accretion disks provide a sensitive and versatile probe of strong-gravity environments and offer a valuable framework for exploring deviations from the Kerr geometry in a model-independent manner. Future extensions of this work may include magnetized disks, non-thermal emission processes, or direct comparisons with observational data to further assess the detectability of effective higher-curvature signatures.


\begin{thebibliography}{99}

\bibitem{1}
Carroll, S. M. (2019). Spacetime and geometry. Cambridge University Press.

\bibitem{2}
K. Schwarzschild, Über das Gravitationsfeld eines Massenpunktes nach der Einsteinschen Theorie, Sitzungsberichte der Königlich Preußischen Akademie der Wissenschaften (Berlin), 189–196 (1916).

\bibitem{3}
Kerr, R. P. (1963). Gravitational field of a spinning mass as an example of algebraically special metrics. Physical review letters, 11(5), 237

\bibitem{4}
Baumann, D. (2022). Cosmology. Cambridge University Press.

\bibitem{5}
Clifton, T., Ferreira, P. G., Padilla, A., \& Skordis, C. (2012). Modified gravity and cosmology. Physics reports, 513(1-3), 1-189.

\bibitem{6}
Zwiebach, B. (1985). Curvature squared terms and string theories. Physics Letters B, 156(5-6), 315-317.

\bibitem{7}
Glavan, D., \& Lin, C. (2020). Einstein-Gauss-Bonnet gravity in four-dimensional spacetime. Physical review letters, 124(8), 081301.

\bibitem{08}
Aoki, K., Gorji, M. A., \& Mukohyama, S. (2020). A consistent theory of D → 4 Einstein-Gauss-Bonnet gravity. Physics Letters B, 810, 135843.

\bibitem{008}
Hennigar, R. A., Kubizňák, D., Mann, R. B., \& Pollack, C. (2020). On taking the D → 4 limit of Gauss-Bonnet gravity: theory and solutions. Journal of High Energy Physics, 2020(7), 1-18

\bibitem{8}
Murodov, S., Rayimbaev, J., Ahmedov, B., \& Karimbaev, E. (2023). Quasiperiodic oscillations and dynamics of test particles around quasi-and non-Schwarzschild black holes. Universe, 9(9), 391.

\bibitem{9}
Ai, W. Y. (2020). A note on the novel 4D Einstein–Gauss–Bonnet gravity. Communications in Theoretical Physics, 72(9), 095402.

\bibitem{10}
M. Heydari-Fard, M. F. da Silva, N. R. da Silva, H. Razmi, Charged black holes in 4D Einstein-Gauss-Bonnet gravity: exact solutions and thermodynamics, Gen. Rel. Grav. 54, 124 (2022). arXiv:2204.04927

\bibitem{11}
Konoplya, R. A., \& Zhidenko, A. (2020). Black holes in the four-dimensional Einstein-Lovelock gravity. Physical Review D, 101(8), 084038.

\bibitem{12}
Kumar, R., \& Ghosh, S. G. (2020). Rotating black holes in 4D Einstein-Gauss-Bonnet gravity and its shadow. Journal of Cosmology and Astroparticle Physics, 2020(07), 053.

\bibitem{13}
Event Horizon Telescope Collaboration. (2019). First M87 event horizon telescope results. I. The shadow of the supermassive black hole. arXiv preprint arXiv:1906.11238.

\bibitem{14}
Fernandes, P. G., Carrilho, P., Clifton, T., \& Mulryne, D. J. (2022). The 4D Einstein–Gauss–Bonnet theory of gravity: a review. Classical and Quantum Gravity, 39(6), 063001.

\bibitem{15}
Zhang, C. Y., Li, P. C., \& Guo, M. (2020). Greybody factor and power spectra of the Hawking radiation in the 4 D Einstein–Gauss–Bonnet de-Sitter gravity. The European Physical Journal C, 80(9), 874.

\bibitem{010}
Bolokhov, S. V., \& Skvortsova, M. (2025). Review of analytic results on quasinormal modes of black holes. arXiv preprint arXiv:2504.05014.

\bibitem{16}
Nojiri, S. S. D. O., Odintsov, S. D., \& Oikonomou, V. (2017). Modified gravity theories on a nutshell: Inflation, bounce and late-time evolution. Physics Reports, 692, 1-104.

\bibitem{17}
Lovelock, D. (1971). The Einstein tensor and its generalizations. Journal of Mathematical Physics, 12(3), 498-501.

\bibitem{18}
Gürses, M., Şişman, T. Ç., \& Tekin, B. (2020). Is there a novel Einstein–Gauss–Bonnet theory in four dimensions?. The European Physical Journal C, 80(7), 647.

\bibitem{19}
Bonifacio, J., Hinterbichler, K., \& Johnson, L. A. (2020). Amplitudes and 4D Gauss-Bonnet Theory. Physical Review D, 102(2), 024029.

\bibitem{20}
Hennigar, R. A., Jahani Poshteh, M. B., \& Mann, R. B. (2018). Shadows, signals, and stability in Einsteinian cubic gravity. Physical Review D, 97(6), 064041.

\bibitem{21}
Fernandes, P. G. (2020). Charged black holes in AdS spaces in 4D Einstein Gauss-Bonnet gravity. Physics Letters B, 805, 135468.

\bibitem{012}
Meng, L., Xu, Z., \& Tang, M. (2025). Periodic Orbits and Gravitational Wave Radiation of Black Holes in 4D-EGB gravity. arXiv preprint arXiv:2506.05015.

\bibitem{40}
Frank, J., King, A., \& Raine, D. J. (2002). Accretion Power in Astrophysics (3rd ed.). Cambridge University Press.

\bibitem{69}
Hoyle, F., \& Lyttleton, R. A. (1939, July). The effect of interstellar matter on climatic variation. In Mathematical proceedings of the Cambridge philosophical society (Vol. 35, No. 3, pp. 405-415). Cambridge University Press.

\bibitem{70}
Prendergast, K. H., \& Burbidge, G. R. (1989). On the nature of some galactic X-ray sources. The Astrophysical Journal, 151, L83.

\bibitem{22}
Shakura, N. I., \& Sunyaev, R. A. (1973). Black holes in binary systems. Observational appearance. Astronomy and Astrophysics, Vol. 24, p. 337-355, 24, 337-355.

\bibitem{44}
Page, D. N., \& Thorne, K. S. (1974). Disk-accretion onto a black hole. Time-averaged structure of accretion disk. The Astrophysical Journal, 191, 499-506. 

\bibitem{23}
Abramowicz, M. A., \& Fragile, P. C. (2013). Foundations of black hole accretion disk theory. Living Reviews in Relativity, 16(1), 1.

\bibitem{71}
Remillard, R. A., \& McClintock, J. E. (2006). X-ray properties of black-hole binaries. Annu. Rev. Astron. Astrophys., 44(1), 49-92.

\bibitem{72}
Bambi, C., Cárdenas-Avendaño, A., Dauser, T., García, J. A., \& Nampalliwar, S. (2017). Testing the Kerr black hole hypothesis using X-ray reflection spectroscopy. The Astrophysical Journal, 842(2), 76.

\bibitem{73}
McClintock, J. E., Narayan, R., \& Steiner, J. F. (2014). Black hole spin via continuum fitting and the role of spin in powering transient jets. Space Science Reviews, 183(1), 295-322.

\bibitem{74}
Kormendy, J., \& Ho, L. C. (2013). Coevolution (or not) of supermassive black holes and host galaxies. Annual Review of Astronomy and Astrophysics, 51(1), 511-653.

\bibitem{75}
Balbus, S. A., \& Hawley, J. F. (1998). Instability, turbulence, and enhanced transport in accretion disks. Reviews of modern physics, 70(1), 1.

\bibitem{76}
Latter, H. N., Lesaffre, P., \& Balbus, S. A. (2009). MRI channel flows and their parasites. Monthly Notices of the Royal Astronomical Society, 394(2), 715-729.

\bibitem{77}
McKinney, J. C., Tchekhovskoy, A., \& Blandford, R. D. (2012). General relativistic magnetohydrodynamic simulations of magnetically choked accretion flows around black holes. Monthly Notices of the Royal Astronomical Society, 423(4), 3083-3117.

\bibitem{78}
Di Matteo, T., Springel, V., \& Hernquist, L. (2005). Energy input from quasars regulates the growth and activity of black holes and their host galaxies. nature, 433(7026), 604-607.

\bibitem{111} Mustafa, G., Donmez, O., Gogoi, D. J., Ghosh, S. G., Hussain, I.,  Yuan, C. (2026). Dynamics, Ringdown, and Accretion-Driven Multiple Quasi-Periodic Oscillations of Kerr-Bertotti-Robinson Black Holes. arXiv preprint arXiv:2602.08911.

\bibitem{112} Mustafa, G., Donmez, O., Errehymy, A., Javed, F., Ditta, A., Naseer, T., ...  Atamurotov, F. (2025). Analytical and numerical study of accretion processes around charged spherically symmetric black holes in scalar-tensor Gauss-Bonnet gravity. arXiv preprint arXiv:2511.12670.

\bibitem{113} He, G., Huang, J., Feng, Z., Mustafa, G.,  Lin, W. (2026). Strong field gravitational lensing of particles by a black-bounce-Schwarzschild black hole. arXiv preprint arXiv:2602.04218.
\bibitem{114} Donmez, O., Ghosh, S. G., Yousaf, M., Mustafa, G.,  Atamurotov, F. (2026). Accretion flow around Kerr metric in the infra-red limit of asymptotically safe gravity. arXiv preprint arXiv:2601.14113.
\bibitem{115} Sucu, E., Sakallı, I., Donmez, O.,  Mustafa, G. (2025). Confining nonlinear electrodynamics black holes: from thermodynamic phases to high-frequency phenomena with accretion process. arXiv preprint arXiv:2512.19448.
\bibitem{116} Donmez, O., Mustafa, G., Yousaf, M., Javed, F., Saidov, I.,  Atamurotov, F. (2025). Origin of Quasi-Periodic Oscillations and Accretion Process in X-Ray Binaries around Quantum Lee-Wick Black Hole. arXiv preprint arXiv:2512.17358.
\bibitem{117} Donmez, O., Mustafa, G., Chaudhary, H., Yousaf, M., Bouzenada, A., Ditta, A., Atamurotov, F. (2026). Relativistic accretion process onto rotating black holes in Einstein-Euler-Heisenberg nonlinear electrodynamic gravity. Physics of the Dark Universe, 102271.
\bibitem{118} Mustafa, G., Javed, F., Maurya, S. K., Ditta, A., Donmez, O., Naseer, T., ...  Atamurotov, F. (2025). Magnetized particle motion and accretion process with shock cone morphology around a decoupled hairy black holes. arXiv preprint arXiv:2511.17137.


\bibitem{79}
Fabian, A. C. (2012). Observational evidence of active galactic nuclei feedback. Annual Review of Astronomy and Astrophysics, 50(1), 455-489.

\bibitem{80}
Abuter, R., Amorim, A., Bauböck, M., Berger, J. P., Bonnet, H., Brandner, W., ... \& Yazici, S. (2018). Detection of orbital motions near the last stable circular orbit of the massive black hole SgrA. Astronomy \& Astrophysics, 618, L10.

\bibitem{90}
Nozari, K., Saghafi, S., \& Aliyan, F. (2025). Investigating QED effects on the thin accretion disk properties around rotating Euler–Heisenberg black holes. The European Physical Journal C, 85(7), 735.

\bibitem{91}
Jusufi, K., Anand, A., Saghafi, S., Cuadros-Melgar, B., \& Nozari, K. (2025). Black holes surrounded by massive vector fields in Kaluza–Klein gravity. The European Physical Journal C, 85(5), 528.

\bibitem{92}
Nozari, K., Saghafi, S., Hajebrahimi, M., \& Jusufi, K. (2025). Circular orbits and accretion disk around a deformed-Schwarzschild black hole in loop quantum gravity. arXiv preprint arXiv:2504.12486.

\bibitem{93}
Nozari, K., Saghafi, S., \& Hassani, M. (2025). Accretion onto a charged black hole in consistent 4D Einstein-Gauss-Bonnet gravity. Journal of High Energy Astrophysics, 45, 214-230.

\bibitem{94}
Salahshoor, K., \& Nozari, K. (2018). Circular orbits and accretion process in a class of Horndeski/Galileon black holes. The European Physical Journal C, 78(6), 486.

\bibitem{95}
Faraji, S. (2025). The impact of compact object deformation on thin accretion disk properties. The European Physical Journal C, 85(2), 148.

\bibitem{96}
Kobialko, K., Gal'tsov, D., \& Molchanov, A. (2025). Gravitational shadow and emission spectrum of thin accretion disks in a plasma medium. arXiv preprint arXiv:2505.07993.

\bibitem{97}
Ouyang, Y., Zhou, X., Chen, S., \& Jing, J. (2025). Thin accretion disk around a Kerr black hole immersed in swirling universes. arXiv preprint arXiv:2504.20582.

\bibitem{98}
Capozziello, S., Gambino, S., \& Luongo, O. (2025). Comparing Bondi and Novikov-Thorne accretion disk luminosity around regular black holes. Physics of the Dark Universe, 101950.

\bibitem{99}
Li, Z., \& Guo, X. K. (2025). Thin accretion disk around rotating hairy black hole: radiative property and optical appearance. The European Physical Journal C, 85(6), 679.

\bibitem{100}
Zheng, H. B., Wu, M. Q., Li, G. P., \& Jiang, Q. Q. (2025). Shadows and accretion disk images of charged rotating black hole in modified gravity theory. The European Physical Journal C, 85(1), 46.

\bibitem{101}
Ashoorioon, A., Poshteh, M. B. J., \& Luongo, O. (2024). Particle motion and accretion disk around rotating accelerating black holes. arXiv preprint arXiv:2408.07762.

\bibitem{102}
Alloqulov, M., \& Shaymatov, S. (2024). Electric Penrose process and the accretion disk around a 4D-charged Einstein-Gauss-Bonnet black hole. The European Physical Journal Plus, 139(8), 731.

\bibitem{103}
Mustafa, G., Maurya, S. K., Ditta, A., Ray, S., \& Atamurotov, F. (2024). Circular orbits and accretion disk around AdS black holes surrounded by dark fluid with Chaplygin-like equation of state. The European Physical Journal C, 84(7), 690.

\bibitem{104}
Abbas, G., Rehman, H., Zhu, T., Wu, Q., \& Mustafa, G. (2023). Accretion disk around Reissner-Nordstrom black hole coupled with a nonlinear electrodynamics field. arXiv preprint arXiv:2310.04053.

\bibitem{105}
Zanotti, O., \& Pugliese, D. (2015). Von Zeipel’s theorem for a magnetized circular flow around a compact object. General Relativity and Gravitation, 47(4), 44.

\bibitem{023}
Donmez, O. (2024). Perturbing the Stable Accretion Disk in Kerr and 4D Einstein–Gauss–Bonnet Gravities: Comprehensive Analysis of Instabilities and Dynamics. Research in Astronomy and Astrophysics, 24(8), 085001.

\bibitem{24}
Wei, S. W., \& Liu, Y. X. (2021). Testing the nature of Gauss–Bonnet gravity by four-dimensional rotating black hole shadow. The European Physical Journal Plus, 136(4), 436.

\bibitem{024}
Bhar, P., Govender, M., \& Sharma, R. (2017). A comparative study between EGB gravity and GTR by modeling compact stars. The European Physical Journal C, 77(2), 109.

\bibitem{25}
Guo, M., \& Li, P. C. (2020). Innermost stable circular orbit and shadow of the 4D Einstein–Gauss–Bonnet black hole. The European Physical Journal C, 80(6), 1-8.

\bibitem{26}
Török, G., Abramowicz, M. A., Kluźniak, W., \& Stuchlík, Z. (2005). The orbital resonance model for twin peak kHz quasi periodic oscillations in microquasars. Astronomy \& Astrophysics, 436(1), 1-8.

\bibitem{27}
Newman, E. T., Couch, E., Chinnapared, K., Exton, A., Prakash, A., \& Torrence, R. (1965). Metric of a rotating, charged mass. Journal of mathematical physics, 6(6), 918-919.

\bibitem{28}
Everitt, C. F., DeBra, D. B., Parkinson, B. W., Turneaure, J. P., Conklin, J. W., Heifetz, M. I., ... \& Wang, S. (2011). Gravity probe B: final results of a space experiment to test general relativity. Physical Review Letters, 106(22), 221101.

\bibitem{29}
Carter, B. (1968). Global structure of the Kerr family of gravitational fields. Physical Review, 174(5), 1559.

\bibitem{30}
Von Zeipel, H. (1924). The radiative equilibrium of a rotating system of gaseous masses. Monthly Notices of the Royal Astronomical Society, Vol. 84, p. 665-683, 84, 665-683.

\bibitem{31}
Ghosh, S. G., \& Maharaj, S. D. (2020). Radiating black holes in the novel 4D Einstein–Gauss–Bonnet gravity. Physics of the Dark Universe, 30, 100687.

\bibitem{82}
Chandrasekhar, S. (1933). The equilibrium of distorted polytropes. I. The rotational problem. Monthly Notices of the Royal Astronomical Society, Vol. 93, p. 390-406, 93, 390-406.

\bibitem{83}
Collins, G. W. (1978). The virial theorem in stellar astrophysics. Tucson, Ariz., Pachart Publishing House (Astronomy and Astrophysics Series. Volume 7), 1978. 143 p.

\bibitem{84}
Lara, F. E., \& Rieutord, M. (2011). Gravity darkening in rotating stars. Astronomy \& Astrophysics, 533, A43.

\bibitem{81}
Tassoul, J. L. (2015). Theory of Rotating Stars.(PSA-1), Volume 1. Princeton University Press.

\bibitem{85}
Townsend, R. H. D. (1997). Spectroscopic modelling of non-radial pulsation in rotating early-type stars. Monthly Notices of the Royal Astronomical Society, 284(4), 839-858.

\bibitem{86}
Lara, F. E., \& Rieutord, M. (2013). Self-consistent 2D models of fast-rotating early-type stars. Astronomy \& Astrophysics, 552, A35.

\bibitem{87}
Wade, G. A., Grunhut, J. H., \& MiMeS Collaboration. (2012). The MiMeS survey of magnetism in massive stars. arXiv preprint arXiv:1206.5163.

\bibitem{88}
de Souza, A. D., Kervella, P., Jankov, S., Abe, L., Vakili, F., Di Folco, E., \& Paresce, F. (2003). The spinning-top Be star Achernar from VLTI-VINCI. Astronomy \& Astrophysics, 407(3), L47-L50.

\bibitem{89}
Collaboration, G., Mignard, F., Klioner, S. A., Lindegren, L., Hernández, J., Bastian, U., ... \& Hodgkin, S. T. (2018). Gaia Data Release 2. A\&A, 616, A14.

\bibitem{106}
Abramowicz, M. A. (1971). The Relativistic von Zeipel's Theorem. Acta Astronomica, Vol. 21, p. 81, 21, 81.

\bibitem{109}
Newman, E. T., \& Janis, A. I. (1965). Note on the Kerr spinning‐particle metric. Journal of Mathematical Physics, 6(6), 915-917.

\bibitem{110}
Newman, E. T., Couch, E., Chinnapared, K., Exton, A., Prakash, A., \& Torrence, R. (1965). Metric of a rotating, charged mass. Journal of mathematical physics, 6(6), 918-919.

\bibitem{47}
Papnoi, U., \& Atamurotov, F. (2022). Rotating charged black hole in 4D Einstein–Gauss–Bonnet gravity: Photon motion and its shadow. Physics of the Dark Universe, 35, 100916.

\bibitem{50}
Thirring, H. (1918). Über die Wirkung rotierender ferner Massen in der Einsteinschen Gravitationstheorie. Physikalische Zeitschrift, 19, 33.

\bibitem{51}
Thirring, H. (1921). Berichtigung zu meiner Arbeit:" Über die Wirkung rotierender Massen in der Einsteinschen Gravitationstheorie". Physikalische Zeitschrift, 22, 29.

\bibitem{52}
Lense, J., \& Thirring, H. (1918). Über den Einfluß der Eigenrotation der Zentralkörper auf die Bewegung der Planeten und Monde nach der Einsteinschen Gravitationstheorie. Physikalische zeitschrift, 19, 156.

\bibitem{48}
Cohen, J. M. (1967). Dragging of inertial frames by rotating masses. Relativity Theory and Astrophysics. Vol. 1: Relativity and Cosmology, 8, 200.

\bibitem{108}
Dugas, R. (2012). A history of mechanics. Courier Corporation.

\bibitem{53}
Pugh, G. E. (2003). Proposal for a satellite test of the Coriolis predictions of General Relativity. In Nonlinear Gravitodynamics: The Lense-Thirring Effect (pp. 414-426).

\bibitem{49}
Costa, L. F. O., \& Natário, J. (2021). Frame-dragging: meaning, myths, and misconceptions. Universe, 7(10), 388.

\bibitem{54}
Schiff, L. I. (1960). Possible new experimental test of general relativity theory. Physical Review Letters, 4(5), 215.

\bibitem{55}
Rindler, W. (1997). The case against space dragging. Physics Letters A, 233(1-2), 25-29.

\bibitem{58}
Heydari-Fard, M., Heydari-Fard, M., \& Sepangi, H. R. (2021). Thin accretion disks around rotating black holes in 4 D Einstein–Gauss–Bonnet gravity. The European Physical Journal C, 81(5), 473.

\bibitem{59}
Liu, C., Zhu, T., \& Wu, Q. (2021). Thin accretion disk around a four-dimensional Einstein-Gauss-Bonnet black hole. Chinese Physics C, 45(1), 015105.

\bibitem{56}
Kurmanov, E., Boshkayev, K., Giambò, R., Konysbayev, T., Luongo, O., Malafarina, D., \& Quevedo, H. (2022). Accretion disk luminosity for black holes surrounded by dark matter with anisotropic pressure. The Astrophysical Journal, 925(2), 210.

\bibitem{57}
Alloqulov, M., Shaymatov, S., Ahmedov, B., \& Jawad, A. (2024). Radiation properties of the accretion disk around a black hole in Einstein-Maxwell-scalar theory. Chinese Physics C, 48(2), 025101.

\bibitem{32}
Frank, J., King, A., \& Raine, D. J. (2002). Accretion Power in Astrophysics (3rd ed.). Cambridge University Press.

\bibitem{33}
Kato, S., Fukue, J., \& Mineshige, S. (2008). Black-Hole Accretion Disks: Towards a New Paradigm. Kyoto University Press.

\bibitem{34}
Shakura, N. I., \& Sunyaev, R. A. (1973). Black holes in binary systems. Observational appearance. Astronomy and Astrophysics, 24, 337-355.

\bibitem{35}
Kerr, R. P. (1963). Gravitational field of a spinning mass as an example of algebraically special metrics. Physical Review Letters, 11(5), 237-238.

\bibitem{36}
Bardeen, J. M., Press, W. H., \& Teukolsky, S. A. (1972). Rotating black holes: locally nonrotating frames, energy extraction, and scalar synchrotron radiation. The Astrophysical Journal, 178, 347-370.

\bibitem{37}
Li, L. -X., Zimmerman, E. R., Narayan, R., \& McClintock, J. E. (2005). Multitemperature blackbody spectrum of a thin accretion disk around a Kerr black hole: Model computations and comparison with observations. The Astrophysical Journal Supplement Series, 157(2), 335-370.

\bibitem{38}
Koide, S., \& Baba, T. (2014). Electromagnetic energy extraction from a Kerr-Newman black hole in a magnetic field. The Astrophysical Journal, 792(1), 88.

\bibitem{39}
Bamba, K., \& Odintsov, S. D. (2010). Inflation and late-time cosmic acceleration in non-minimal Maxwell-F(R) gravity and the generation of large-scale magnetic fields. Journal of Cosmology and Astroparticle Physics, 2010(04), 008.

\bibitem{41}
Rybicki, G. B., \& Lightman, A. P. (1986). Radiative Processes in Astrophysics. Wiley-VCH. 

\bibitem{42}
Mitsuda, K., et al. (1984). Energy spectra of low-mass binary X-ray sources observed from Tenma. Publications of the Astronomical Society of Japan, 36, 741-759. 

\bibitem{43}
Davis, S. W., \& El-Abd, S. (2019). The X-ray spectra of accretion discs in active galactic nuclei. Monthly Notices of the Royal Astronomical Society, 486(1), 1223-1231. 

\bibitem{60}
Page, D. N., \& Thorne, K. S. (1974). Disk-accretion onto a black hole. Time-averaged structure of accretion disk. Astrophysical Journal, Vol. 191, pp. 499-506 (1974), 191, 499-506.

\bibitem{61}
Li, L. X., Zimmerman, E. R., Narayan, R., \& McClintock, J. E. (2005). Multitemperature blackbody spectrum of a thin accretion disk around a Kerr black hole: model computations and comparison with observations. The Astrophysical Journal Supplement Series, 157(2), 335.

\bibitem{45}
Li, L. -X., Zimmerman, E. R., Narayan, R., \& McClintock, J. E. (2005). Multitemperature blackbody spectrum of a thin accretion disk around a Kerr black hole: Model computations and comparison with observations. The Astrophysical Journal Supplement Series, 157(2), 335-370. 

\bibitem{62}
Ross, R. R., Fabian, A. C., \& Mineshige, S. (1992). The spectra of accretion discs in active galactic nuclei. Monthly Notices of the Royal Astronomical Society, 258(1), 189-197.

\bibitem{46}
Peterson, B. M. (1997). An Introduction to Active Galactic Nuclei. Cambridge University Press.

\bibitem{63}
Rybicki, G. B., \& Lightman, A. P. (2024). Radiative processes in astrophysics. John Wiley \& Sons.

\bibitem{107}
Abramowicz, M. A., Miller, J. C., \& Stuchlík, Z. (1993). Concept of radius of gyration in general relativity. Physical Review D, 47(4), 1440.

\bibitem{64}
Zanotti, O., \& Pugliese, D. (2015). Von Zeipel’s theorem for a magnetized circular flow around a compact object. General Relativity and Gravitation, 47(4), 44.

\bibitem{65}
Jefremov, P. I., \& Perlick, V. (2016). Circular motion in NUT space-time. Classical and Quantum Gravity, 33(24), 245014.

\bibitem{66}
Abramowicz, M. A. (1990). Centrifugal force-a few surprises. Monthly Notices of the Royal Astronomical Society, Vol. 245, NO. 4/AUG15, P. 733, 1990, 245, 733.

\bibitem{67}
Chakrabarti, S. K. (1990). Von-Zeipel Surfaces. Monthly Notices of the Royal Astronomical Society, Vol. 245, NO. 4/AUG15, P. 747, 1990, 245, 747.

\bibitem{68}
Bini, D., de Felice, F., \& Jantzen, R. T. (2003). Centripetal acceleration and centrifugal force in general relativity. In Nonlinear Gravitodynamics: The Lense-Thirring Effect (pp. 119-127).

\end{thebibliography}
\end{document}